\documentclass[a4paper,11pt]{article}
\pdfoutput=1
\usepackage{jheppub}

\usepackage{amsmath,amssymb,amsfonts,cases}
\usepackage{amsthm}
\usepackage{graphicx,caption,subcaption}
\usepackage{dsfont}
\usepackage{slashed}
\usepackage{wrapfig}
\usepackage{mathtools}
\usepackage{tikz}

\usepackage{tikz-cd}
\usepackage{physics}
\usepackage{hyperref}
\usepackage{simpler-wick}
\usepackage[inline]{enumitem}

\def\be{\begin{equation}}
\def\ee{\end{equation}}

\def\bal{\begin{equation}\begin{aligned}}
\def\eal{\end{aligned}\end{equation}}

\def\d{{\rm d}}

\def\comment#1{}

\title{\LARGE Bootstrapping Boundary QED Part I}
\author{Samuel Bartlett-Tisdall,}
\author{Christopher P.\ Herzog,}
\author{Vladimir Schaub}
\affiliation{Department of Mathematics, King's College London, \\  Strand, London, WC2R 2LS, UK}

\emailAdd{samuel.c.bartlett-tisdall@kcl.ac.uk}
\emailAdd{christopher.herzog@kcl.ac.uk}
\emailAdd{vladimir.schaub@kcl.ac.uk}

\abstract{ 
We use the numerical conformal bootstrap to study boundary quantum electrodynamics, the theory of a four dimensional photon in a half space coupled to charged conformal matter on the boundary.  This system
is believed to be a boundary conformal field theory with an exactly 
marginal coupling corresponding to the strength of the interaction 
between the photon and the matter degrees of freedom.  In part one
of this project, we present three results.  We show how 
the Maxwell equations put severe constraints on boundary three-point functions
involving two currents and a symmetric traceless tensor.  
We use semi-definite programming to show that any three dimensional 
conformal field theory with a global U(1) symmetry must have a spin two
gap less than about 1.05.  Finally, combining a numerical bound on an OPE
coefficient and some Ward identities involving the current and the displacement
operator, we bound the displacement operator two-point function above.
This upper bound also 
constrains a boundary contribution to the anomaly in the
trace of the stress tensor for these types of theories.

}

\makeatletter
\def\@fpheader{\vspace{0cm}}
\makeatother

\begin{document}
\maketitle

\section{Introduction}

Four dimensional quantum electrodynamics (three spatial dimensions and one time) is an important example of quantum field theory, underpinning our understanding of the interaction between light and charged particles in the world around us.  Three dimensional QED has had a more limited experimental impact but nonetheless has featured prominently in theoretical progress over the years.  For example, by tuning the number of electron flavors $N_f$, the low energy limit of this theory passes from a conformal fixed point at large $N_f$ to a theory with a mass gap at small $N_f$, providing a model of confinement \cite{Pisarski:1984dj,Appelquist:1988sr,Nash:1989xx}.  This work is about a mixed version of these two theories, that we will call boundary QED, where the photon is four dimensional but the charged matter is confined to a 2+1 dimensional plane.  In a sense that we will make more precise, the model is like graphene.  Less well studied than its pure three dimensional and four dimensional cousins, it has inspired a small but fruitful line of inquiry over the years, with early references here \cite{Marino:1992xi,Gonzalez:1993uz,Gorbar:2001qt}.

As argued by refs.\ \cite{Herzog:2017vj,Dudal:2018mms,DiPietro:2019hqe}, boundary QED 
is a boundary conformal field theory (bCFT). 
The main purpose of this work is the numerical conformal
bootstrap \cite{Poland:2018epd} of boundary QED.  
While numerical bootstrap of bulk correlation functions
in bCFT face positivity issues \cite{Liendo:2013tw},
exceptionally there is a way around this problem for boundary QED.
The strategy we employ here was inspired by previous work \cite{Lauria:2021ut,Behan:2020nsf,Behan:2021tcn} 
where instead of a free photon, a free scalar is coupled to degrees of freedom on the boundary. The two ingredients are a restriction of the boundary spectrum by the bulk free equations of motions, and a locality constraint on bulk-reconstructed operators, which puts severe constraints on the boundary correlation functions.  It is these constraints ultimately that we hope to impose in the numerical bootstrap.

In our case, we find that the Maxwell equations
imply that there are exactly two currents $J_{(1)}$ and $J_{(2)}$ in the boundary operator expansion (BOE) of the bulk Maxwell field $F_{\mu\nu}$. Of these currents, one has even and one has odd spatial parity, and as a result, we often refer to these as an electric current $E$ and a magnetic current $B$. Ultimately, we will be using our constraints in the numerical conformal bootstrap of
$\langle J_{(a_1)} (x_1) J_{(a_2)}(x_2) J_{(a_3)}(x_3) J_{(a_4)}(x_4) \rangle$. Before we get there, however, 
we first need to provide
a more detailed description of boundary QED and also a discussion of its two and three-point correlation functions. In section \ref{sec:perturbativeperspective}, we start with a perturbative approach to the theory, to set up the problem and define the objects of interest. Of particular relevance to the numerical bootstrap, we review how the gauge field coupling gives rise to an anomalous dimension of the boundary stress tensor. This allows us to use the dimension of the leading boundary spin two operator as a substitute for the coupling in our numerical analysis. We also discuss the displacement operator,
an operator with protected scaling dimension that every bCFT must contain in its spectrum.

In section \ref{sec:FF}, we move on to a discussion of two-point functions.  We compute the BOE of a free Maxwell field $F_{\mu\nu}$ by using a Taylor series expansion into the bulk. The result can be expressed entirely in terms of the currents $J_{(a)}$ and their derivatives. Using the BOE, we compute all the two-point functions between $F_{\mu\nu}$ and the $J_{(a)}$.  These two-point functions are informed by the symmetries of the theory and tell us about the moduli
space of possible boundary conditions of the Maxwell field. 
By imposing unitarity, we find that this moduli space
 can be parametrized by a unit disk cross the half line, and discuss its physical significance.
In particular, 
the $\langle J_{(a)}(x) J_{(b)}(y) \rangle$ two-point functions determine the conductive response of the system
to an oscillating electric field.   To the extent that boundary QED is like graphene, some constraints
on these two-point functions may help resolve a longstanding puzzle about the optical conductivity of graphene,
as we discuss in more detail in the Discussion in section \ref{sec:discussion}.

After discussing the two-point functions, we move to three-point functions in section \ref{sec:threepoint}. For the numerical bootstrap of the correlation function of four currents 
\[
\langle J_{(a_1)} (x_1) J_{(a_2)}(x_2) J_{(a_3)}(x_3) J_{(a_4)}(x_4) \rangle \ ,
\]
 we need to compute three-point functions of the form $\langle J_{(a)}(x_1) J_{(b)}(x_2) T_\ell(x) \rangle$. Here, $T_\ell$ is a symmetric traceless tensor of spin $\ell$ that can appear in the OPE of two of the currents $J_{(a)}$. 
Similar to what happens for a free in the bulk scalar  \cite{Lauria:2021ut,Behan:2020nsf,Behan:2021tcn},\footnote{%
  These constraints have since been generalized for interacting scalar and stress-tensor fields in AdS \cite{Levine:2023ywq,Meineri:2023mps}, where regulating the bulk analytic structure implies a sum rule on the
  boundary OPE decomposition.  Free theories provide particularly simple example of these more general sum rules.
}
fixing $T_{\ell}$, we find that the data for $\langle J_{(a)}(x_1)J_{(b)}(x_2) T_\ell(x) \rangle$ is generically determined by 
the data for any single one of them, for example $\langle J_{(1)}(x_1) J_{(1)}(x_2) T_\ell(x) \rangle$.
The relation comes from a regularity constraint on the bulk-boundary-boundary correlation function
$\langle F_{\mu\nu}(x_1, x_\perp) J_{(a)}(x_2) T_\ell(x) \rangle$.  
Consider the case $T_0$. The three-point functions $\langle J_{(a)}(x_1) J_{(b)}(x_2) T_0(x) \rangle$ are fixed by conformal symmetry up to constants which we call $\gamma_{ab}$ and $\widetilde \gamma_{ab}$, where the tilde denotes odd parity. The new result is then the relation
between these constants:
\begin{equation}
\label{gammapreview}
\widetilde \gamma_{ab}= \frac{(\Delta-1)(\Delta-3)}{(\Delta-2)\Delta} \tan \left(\frac{\pi \Delta}{2} \right) \gamma_{cb}
\, {\epsilon^c}_a \ .
\end{equation}

With a better grasp on the theory, we turn to the numerical conformal bootstrap in section \ref{sec:bootstrap}. 
While we intend ultimately to enforce constraints such as (\ref{gammapreview}) between the OPE coefficients,  
for this paper 
we restrict the bootstrap to four-point functions of a single current, which in this boundary QED context could be either
$\langle E(x_1) E(x_2) E(x_3) E(x_4) \rangle$ or $\langle B(x_1) B(x_2) B(x_3) B(x_4) \rangle$.
Of course, without constraints, we are looking at any 3d CFT with a global U(1) symmetry.  In order
to refine the search space, we focus on two special operators, an even parity scalar of dimension four and the lowest
dimension spin two operator.  In any bCFT, we are guaranteed the existence of a displacement operator $D$, which is the operator sourced by the location of the boundary.  
As the operator can be identified with the boundary limit of the normal-normal component of the bulk stress tensor, it must have protected scaling dimension equal to the bulk space-time dimension and even parity (in theories with parity symmetry).
While before coupling the free fermions to the bulk, they have their own stress tensor, once we introduce
$g \neq 0$, energy can leak off the boundary and the divergence of this boundary stress tensor no longer vanishes.
In particular, its dimension must float up from the unitarity bound of three.  These are exactly the two operators which we explore. 

In our bootstrap analysis, we find two results associated with these operators.  
Let the spin $\ell$ gap be defined as the scaling dimension of the smallest spin $\ell$ operator minus the corresponding unitarity bound.  
The first result is that there appear to be no three dimensional CFTs with a spin 2 gap slightly larger than one, 1.05 to be precise.  Thus having coupled any boundary CFT with a U(1) symmetry to the bulk Maxwell field, the operator that used to be the boundary stress tensor can acquire an anomalous dimension that is no larger than about one.
Moreover, we have some indication that once the spin 2 gap is large enough, the only possible CFT is a generalized free vector field (GFVF), which
has a lowest spin two operator of dimension 4.  
(A GFVF is defined to be a CFT constructed from a conserved vector $V_i$ assuming that all the correlation functions follow from Wick's Theorem.)
We speculate that with a more sophisticated numerical bootstrap analysis, for example using a system of mixed correlators, the  bound of 1.05 could be pushed down closer to one.  Indeed, it would be interesting to see if the bound could be proven analytically using extremal functionals, along the lines of
\cite{Caron-Huot:2020adz}.

The second result is more specific to our boundary QED setup.  There is a Ward identity \cite{DiPietro:2019hqe}
which relates $\langle J_{(a)} J_{(b)} D \rangle$ 
to $\langle J_{(a)} J_{(b)} \rangle$ and $\langle D D \rangle$. Note by conformal invariance, all of these correlation functions are completely determined by a handful of constants.  
Using this Ward identity,  we are able to bound the constant $C_D$ 
that determines $\langle DD \rangle$.  
We find (numerically) $C_D \leq 1.728 \, C_D^{free}$ where $C_D^{free}$ is the value for $D = V_\mu V^\mu$ in the
GFVF.  
This result has added significance as $C_D$ determines also certain boundary terms in the stress tensor trace anomaly for bCFTs \cite{Herzog:2017vj,Herzog:2017kkj}.   Thus by bounding the displacement two-point function, we are also bounding a boundary contribution to the trace anomaly.  

Lastly, we include some additional information in our appendices. Appendix \ref{sec:perimeterimpliesfree} proves that a certain class of boundary conditions for the field $F_{\mu\nu}$ must result in a bulk-boundary decoupling. In appendix \ref{sec:numerics}, we discuss the technical details of our numerics. In appendix \ref{sec:free}, we provide a brief overview of the GFVF, the free scalar, and the free fermion in 3d.  Finally, in appendix \ref{sec:Ward}, we review a Ward identity argument constraining the three-point function
of two currents with the displacement operator.

\section{Perturbative Perspective}
\label{sec:perturbativeperspective}

While our approach is ultimately non-perturbative, from a pedagogical standpoint and to fix notation, it is worthwhile to start by discussing  the decoupled case where the photon does not interact with anything on the boundary. Working in 
Euclidean space, truncated to $x^n>0$, consider a free 4d photon $A_\mu$, described by 
\be
\label{action}
S =\int_{x^n>0} \d^4 x \left(  \frac{1}{4g^2} F_{\mu\nu} F^{\mu\nu} + i\frac{  \theta}{16 \pi^2} F^{\mu\nu} \widetilde F_{\mu\nu} \right) \, ,
\ee  
and we insist that this setup preserves the symmetry of a bCFT.
In this notation Greek indices $\mu$, $\nu$ are bulk while lower case Roman indices $i$, $j$ are reserved for the boundary.  The index $n$ is the direction normal to the boundary.  The Maxwell field strength is as usual $F_{\mu\nu} = \partial_\mu A_\nu - \partial_\nu A_\mu$. We define the (Hodge) dual field strength $\widetilde F_{\mu\nu} \equiv \frac{1}{2} \epsilon_{\mu\nu\lambda \rho} F^{\lambda \rho}$.

The theory is free, the path integral is Gaussian, 
and the saddle point given by a solution
to the equations of motion which are just Maxwell's equations.  Because of the boundary,
there is an extra boundary term in a variational analysis
\be
\label{bc}
-\delta A_i \left( \frac{1}{g^2} F^{ni} + i \frac{ \theta}{4 \pi^2} \widetilde F^{ni}\right) \ ,
\ee
which leads to two boundary conditions, both consistent with conformal symmetry. 
From $\delta A_i = 0$, we have the Dirichlet type condition
$F_{ij} = 0 = \widetilde F_{ni}$, often called absolute boundary conditions. This boundary condition does not allow us to couple the system to boundary degrees of freedom. Alternately, we have the more interesting Neumann type condition, often called relative boundary conditions, 
\[
F^{ni} = -i \frac{ \theta g^2 }{4 \pi^2} \widetilde F^{ni}  \ .
\]
Defining the 
complexified gauge coupling
\be
\tau \equiv \frac{\theta}{2\pi} + \frac{2\pi i}{g^2} \ ,
\ee 
the proportionality constant in the boundary condition depends only on the phase of 
$\tau$.
This relative boundary conditions suggest a rephrasing in terms of boundary currents. We define the ``electric'' and ``magnetic'' fields, 
\begin{align}
     E_i &=  F_{ni} \biggr|_{\rm bry} \ , \,  &  B_i &= i\widetilde F_{ni} \biggr|_{\rm bry} \, .
\end{align}
In the absence of boundary interactions, the natural boundary conditions amount to forcing a linear relation between these two fields, halving the degrees of freedom. 

Field insertions of $E$ or $B$ can be rewritten in terms of some underlying unit-normalised current $V$, whose correlation functions are computed by Wick contractions. This theory is sometimes called a generalized free vector field (GFVF). Its field spectrum is made up of single-trace primaries created by normal ordering the product of fields decomposed over parity, spin $\ell$ and conformal dimension $\Delta$\footnote{%
The form of these operators is schematic, meant only to indicate the total number of derivatives, free indices and parity.  The actual primaries will in general be sums of such terms with coefficients determined by the condition that the sum is annihilated by generators of special conformal transformations. 
Specific examples with $\ell=0$ are presented as
(\ref{GFVFops}).
}
\begin{align}
 	[V\wedge V]_{\Delta, \ell} \sim & \,  {:}z^{i}\epsilon_{ijk}V^{j}\Box^{p}(z\cdot \partial)^{q} V^k{:} \ , & \, \Delta &= 4+2p+J , \, \ell = q+1  \ , \\
	 	[V\cdot V]_{\Delta, \ell} \sim & \, {:}V^{i}\Box^{p}(z\cdot \partial)^{q} V_i{:} \ , & \, \Delta &= 4+2p+q, \, \ell = q \ , \\
	 	[\partial\cdot V\wedge V]_{\Delta,\ell} \sim & \, {:}\epsilon_{ijk} \partial^{i}V^{j}\Box^{p}(z\cdot \partial)^{q} V_k{:} \ , & \, \Delta &= 5+2p+J, \, \ell = q \ ,
\end{align}
where $p$ and $q$ are non-negative integers and $z^i$ is a polarization vector.

Returning to the free Maxwell field with boundary, this theory contains a
bulk stress-tensor. In the index free formalism, it is given by 
\begin{align}
	T(x,z)=\frac{1}{g^2} z^{\mu}F_{\mu\nu}F_{\rho}{}^{\nu}z^{\rho} \ .
\end{align}
We are in particular interested in its boundary limit, from which we can identify two protected boundary primary operators, the displacement operator and the flux vector,%
\begin{align}
	D &= T^{nn} \biggr|_{\rm bry}= \frac{1}{g^2}\left(E^2 + B^2\right) \ , & P^{i}&= T^{ni}  \biggr|_{\rm bry} =\frac{i}{g^2}\epsilon^{ijk}B_{j}E_{k} \ .
\end{align}
One recognizes the energy density and Poynting vector from electromagnetism.
The operator $P^{i}$ is problematic and must be taken out of the spectrum to ensure the conformal invariance 
is preserved (see e.g.\ \cite{McAvity:1993ul,McAvity:1995zd,Herzog:2017vj,Behan:2020nsf,Herzog:2021spv}). In the free setting, vanishing is automatic because the electric and magnetic fields are parallel. Using Wick contractions, it is straightforward to evaluate the norm of the displacement operator
\begin{align}
\label{eq:DDrel}
	\expval{D(x)D(y)}&
	=
	\frac{6}{\pi^4}\frac{1}{(x-y)^8}=C_D^{free}\frac{1}{(x-y)^8}  \ .
\end{align}
The constant $C_D$ has an interpretation in terms of a boundary conformal anomaly coefficient 
\cite{Bianchi:2015liz,Herzog:2017vj,Herzog:2017kkj}.  It forms a crucial piece of CFT data, which we will use as a probe of the theory as we turn on interactions.

Having fixed the CFT data in the free case, one can introduce deformations. The idea is to pick any 3d CFT possessing a U(1) current $\mathcal{J}$. Gauging this theory through coupling to the bulk $F_{\mu\nu}$, 
one can compute correlation functions in the interacting theory through conformal perturbation theory,
\begin{align}
 S \rightarrow S -  \int_{x^{n}=0} \d^{3} x \left( A_i\mathcal{J}^i+\ldots \right) \ ,
\end{align}
where the ellipsis are possible seagull terms. The non-trivial aspect of this system is that the boundary interaction does not (generally) induce a renormalisation of the bulk coupling $\tau$. 
If we take the particular examples of massless fermions or complex scalars, gauging
the U(1) vector symmetry by coupling to the bulk Maxwell field, 
the relative normalization of the boundary kinetic term and the vertex is fixed
by gauge invariance.  As in regular QED, the only way for the gauge coupling to develop a beta function is through 
wave function renormalization of the photon.  However, the photon kinetic term is a bulk term in the Lagrangian, and there is no way for the ``tail to wag the dog'' -- for the boundary to renormalize bulk quantities.  Thus not only is the theory conformal, but it also has an exactly marginal coupling, the gauge coupling \cite{Herzog:2017vj,Dudal:2018mms,DiPietro:2019hqe}.
Indeed, this argument generalizes to include coupling any free Maxwell field to a boundary CFT with a U(1) global symmetry \cite{DiPietro:2019hqe}, with important caveats.  If the boundary theory already 
has a marginal coupling, then
the addition of the gauge coupling $g$ may lead to nonzero beta functions which cannot be tuned away. Further,
the theory may become unstable if the coupling gets sufficiently large; for example, there could be spontaneous generation of a mass gap.
Assuming the coupling is tuned to be in this conformal window, 
the ensuing interacting conformal manifold is our object of study. 

The existence of this space of boundary conditions is non-trivial. Although conformal manifolds are easy to construct for defect CFTs through symmetry breaking, see e.g.\ \cite{Herzog:2023dop,Drukker:2022pxk,Drukker:2022txy,Trepanier:2023tvb}, our example is genuinely interacting.  Moreover, QED with conformal matter on defects with codimension greater than $1$ is likely trivial \cite{Herzog:2022jqv}, making boundary QED special in this regard.  

Turning on interactions means that the currents $E$ and $B$ no longer satisfy a linear relation.
 For simplicity, we will focus on the parity-preserving case $\theta=0$. 
 The modified variational principle gives $E = g^2 {\mathcal J}$ instead of $E=0$.\footnote{%
 More generally, $2 \pi {\mathcal J} = \frac{2\pi}{g^2} E + \frac{\theta}{2\pi} B$,
 and we can also define $2 \pi {\mathcal I} = B$.  The pair $({\mathcal J}, {\mathcal I})$ 
 transform under SL$(2, {\mathbb Z})$ as a doublet and can be identified with the boundary values of 
 \begin{eqnarray*}
 {\mathcal J}_i &=& \left. \frac{\tau F_{ni}^- - \bar \tau F_{ni}^+}{2 \pi i}\right|_{\rm bry} \ , \; \; \;
 {\mathcal I}_i  = \left.  \frac{F_{ni}^- - F_{ni}^+}{2 \pi i} \right|_{\rm bry} \ .
 \end{eqnarray*}
 where $F_{\mu\nu}^\pm \equiv \frac{1}{2} (F_{\mu\nu} \pm \widetilde F_{\mu\nu})$.  
 }
 
 Current two-point functions are determined up to a constant by conformal symmetry.
 Let $\tau_{\mathcal {JJ}}$, $\tau_{EE}$, and $\tau_{BB}$ be the constants
 associated with the two-point functions of ${\mathcal J}$,  $E$, and $B$ respectively.
 The identification $E = g^2 {\mathcal J}$ along with a sum rule \cite{DiPietro:2019hqe}
 we review in section \ref{sec:FF} allows us to deduce the relations
 \begin{equation}
\begin{aligned}
\tau_{EE} &= g^4 \tau_{\mathcal{JJ}}\ , \; \; \;
\tau_{BB} = \frac{2 g^2}{\pi^2} - g^4 \tau_{\mathcal{JJ}} \ .
\end{aligned}
\end{equation}
While these relations are exact, $\tau_{JJ}$ receives corrections from its decoupled-value, making it a function of the  coupling $\tau$.
To visualize the moduli space of boundary conditions for $F_{\mu\nu}$, we will find it convenient to introduce
the reflection coefficient
\begin{equation}
\chi_e \equiv \frac{\tau_{BB} - \tau_{EE}}{\tau_{BB} + \tau_{EE}} = 1 - \pi^2 g^2 \tau_{\mathcal{JJ}}  \ .
\end{equation}

The current two-point functions are not the only interesting quantities. Some perturbative
results are available for a couple of other quantities
that will be important in this work \cite{DiPietro:2019hqe}.\footnote{%
See also \cite{Teber_2012, Kotikov_2013}  for the particular case of coupling to free massless fermions
on the boundary.
}

The norm of the displacement operator 
(and an associated contribution to the trace anomaly)  changes as a function of coupling, making it an interesting observable of the interacting theory, which we will bound in this work. 
The displacement two-point function is corrected at leading order to 
\begin{equation}
C_D = \frac{6}{\pi^4}\left(1  - \pi^2 g^2 \tau_{\mathcal{JJ}}+ O(g^4)  \right)\ .
\end{equation}

A further observable is the leading primary $T_2$ with spin two and smallest conformal dimension.
When the coupling is zero, this operator can be identified with the boundary stress tensor.
However, as we turn on the interaction, this leading primary acquires an anomalous dimension, which can be expressed in terms of the central charge $c$ of the 3d theory as \cite{DiPietro:2019hqe}
\begin{equation}
\Delta_2 = 3 + \frac{6}{5 \pi^2}  \frac{ \tau_{\mathcal{JJ}}}{ c}g^2 + O(g^4) \ .
\end{equation}
For reference $c$ for the Ising model is $3 / 16 \pi^2$.  This fact motivates us to use the dimension of the leading boundary spin 2 operator as a substitute for the coupling in our numerical analysis. We will find evidence that the dimension of $T_2$ is bounded by that in the GFVF theory, i.e.\ four.  
Indeed, one may expect to find the GFVF in the infinite coupling limit.
In the limit $g \to \infty$, the complexified coupling $\tau$ approaches the real axis, which can also 
 be reached by an SL$(2, {\mathbb Z})$ transformation of the free case.   

Another observable, which is accessible to our numerical setup, is the current-current-displacement three-point function coefficients. In a convention specified in eq.\ (\ref{gammaJJD}),
\begin{eqnarray}
\gamma_{BB} &=& \frac{4 g^2}{\pi^4} (1 - \tau_{\mathcal{JJ}} \pi^2 g^2 + O(g^4) ) \ .
\end{eqnarray}
One can also look at the normalised OPE coefficient, which shows up in the numerical bootstrap, 
\begin{align}
\frac{\gamma_{BB}}{\tau_{BB} \sqrt{C_D}} &= \sqrt{\frac{2}{3}} \left( 1 + O(g^4) \right) \ .
\end{align}

\section{The Two-Point Function of the Maxwell Field}
\label{sec:FF}

The goal of this section is to characterize and constrain the boundary conditions on $F_{\mu\nu}$ in the presence
of charged degress of freedom on the boundary.  
We derive the  bulk-to-boundary operator expansion (BOE) \cite{McAvity:1995zd} of $F_{\mu\nu}$, which we then use to compute
$\langle F_{\mu\nu}(x) F_{\rho \lambda}(y) \rangle$ in terms of a moduli space specified by a triplet of numbers.
These numbers  parametrize ${\mathbb R}^+ \times D$ where $D$ is a disk with unit radius.

The content can largely be found in \cite{DiPietro:2019hqe} although our development and perspective is different. In particular, our discussion of the moduli space and its relation to symmetries is new.

\subsection{Boundary Operator Expansion}
\label{sec:BOE}

To compute the BOE of a free Maxwell field, we begin with a Taylor series expansion of $F_{\mu\nu}$.
The resulting boundary objects are then decomposed in terms of boundary irreducible representations, using the building blocks that are the normal vector $n^{\mu}$ to the boundary, and the boundary induced metric $h_{\mu\nu}=\delta_{\mu\nu}-n_{\mu}n_\nu$,
\bal \label{eq:preope}
	F_{\mu\nu}(x,x_\perp)&=\sum \frac{x_\perp^k}{k!}\partial_n^{k} F_{\mu\nu}(x,0)\\
	&=
	\sum \frac{x_\perp^k}{k!}\partial_n^{k} \bigg( (n_\mu h_{\nu i}-n_\nu h_{\mu i})\underbrace{F^{ni}(x,0)}_{E^{i}(x)}+ih_{\mu i}h_{\nu j}\epsilon^{ijm}\underbrace{\frac{\epsilon_{mls}}{2i}F^{ls}(x,0)}_{B_{m}(x)} \bigg) \ .
\eal
Note that $x_\perp \equiv n \cdot x$.
The $E^i$ and $B^i$ currents of section \ref{sec:perturbativeperspective} have reappeared as primary operators in this boundary decomposition.
We work in Euclidean signature throughout, and the factor of $i$ is inherited from the Wick rotation from Lorentzian to Euclidean. 

The electromagnetic fields are defined on the boundary, and to rewrite \eqref{eq:preope} in terms of a BOE, we must trade the normal derivatives for tangential derivatives. To do so, we use the free Maxwell equations
\begin{align}\label{eq:eom}
&\begin{aligned}
	\partial_\mu F^{\mu\nu}&=0 \, , \\
	 \partial_{[\alpha}F_{\beta \gamma]}&=0\, .
\end{aligned}  &\Longleftrightarrow && \begin{aligned}
	\partial_n E^{i}&=+ i \epsilon^{ijk}\partial_j B_k\, , &  \partial \cdot E &= 0\, , \\
	 \partial_n B^i &=-i \epsilon^{ijk}\partial_j E_k 
	\, ,   &  \partial \cdot B &= 0\, .
\end{aligned}
\end{align}
The divergence conditions on the electric and magnetic fields are precisely the conservation conditions, allowing us to reinterpret $E$ and $B$ as conserved currents.
Using the differential relations \eqref{eq:eom}, we gather 
\begin{align}
	\partial_{n}^{(2k)} E^i &= (-\Box_\parallel)^{k}E^{i} \ , & \partial_{n}^{(2k+1)}E^{i}&=+i(- \Box_\parallel)^{k} \epsilon^{ijl}\partial_j B_l \ ,\\
	\partial_{n}^{(2k)} B^i &= (-\Box_\parallel)^{k}B^{i} \ , & \partial_{n}^{(2k+1)}B^{i}&=-i(-\Box_\parallel)^{k} \epsilon^{ijl}\partial_j E_l \ .
\end{align}
Collecting factors and reshuffling slightly, we go from \eqref{eq:preope} to our BOE
\bal
\label{eq:FBOEprelim}
	F_{\mu\nu}&= \sum_{k=0}^{\infty} (-\Box_\parallel)^{k} \left( \frac{x_\perp^{2k}}{(2k)!} (n_\mu h_{\nu i}-n_\nu h_{\mu i}) E^{i}+\frac{x_\perp^{2k+1}}{(2k+1)!} (h_{\mu i}h_{\nu j}-h_{\nu i}h_{\mu j})\partial^{i} E^{j} \right)\\
	&+i\sum_{k=0}^{\infty}  (-\Box_\parallel)^{k} \left( \frac{x_\perp^{2k}}{(2k)!} \epsilon_{\mu\nu m} B^{m}+\frac{x_\perp^{2k+1}}{(2k+1)!}(n_\mu \epsilon_{\nu j l}-n_\nu \epsilon_{\mu j l})\partial^j B^{l}\right) \ ,
\eal
expressed purely in terms of the primary operators $E$ and $B$, which are conserved currents, and their descendants.  

The expression (\ref{eq:FBOEprelim}) 
 can be massaged into a more palatable form. We define the following two differential operators 
 \begin{align}
 \label{eq:DDidefs}
 	\mathcal{D}(x_\perp,\partial)& \equiv \sum_{k=0}^{\infty} \frac{(-\Box_\parallel x_\perp^2)^k}{(2k)!} \ , & \mathcal{D}^{i} (x_\perp,\partial)& \equiv \sum_{k=0}^{\infty} \frac{(-\Box_\parallel x_\perp^2)^k}{(2k+1)!} x_\perp \partial^i \ .
 \end{align}
Relating the 4d and 3d Levi-Civita tensors through the relation $\epsilon^{\alpha\beta\rho\lambda}n_\lambda =\epsilon^{\alpha\beta\rho}$, the final form of the boundary OPE becomes

 \bal\label{eq:Fope}
 	F_{\mu\nu} 
 	&= \bigg( (n_\mu h_{\nu i}-n_\nu h_{\mu i}) \mathcal{D}(x_\perp,\partial)+ (h_{\mu j} h_{\nu i}-h_{\nu j} h_{\mu i}) \mathcal{D}^{j}(x_\perp,\partial) \bigg)E^i\\
 	&-\frac{i}{2}\epsilon_{\mu\nu}{}^{\alpha\beta}\bigg( (n_\alpha h_{\beta i}-n_\beta h_{\alpha i}) \mathcal{D}(x_\perp,\partial)+ (h_{\alpha j} h_{\beta i}-h_{\beta j} h_{\alpha i}) \mathcal{D}^{j}(x_\perp,\partial) \bigg)B^i \ .
\eal
These structures are divided clearly between the normal and tangential directions. In the way we chose to write things here, it is clear that all parity-odd structures will come from resummations involving an odd number of magnetic fields, and that the different OPE channels are exchanged under Hodge-duality. Parity odd data will be pure imaginary, as it should be.

 \subsection{Two-Point Correlators From the BOE}
 \label{sec:FFredux}
 
 The point of this section is to discuss and compute two-point  functions,
 \begin{align}
 \langle E_i(x) E_j(y) \rangle \ , \; \; \;
  \langle E_i(x) B_j(y) \rangle \ , \; \; \;
   \langle B_i(x) B_j(y) \rangle \ , \; \; \; \\
 \langle F_{\mu\nu} (x) E^i(y) \rangle \ , \; \; \;
 \langle F_{\mu\nu} (x) B^i(y) \rangle \ , \; \; \;
 \langle F_{\mu\nu} (x) F_{\lambda \rho}(y) \rangle \ .
 \end{align}
 We will  exploit the BOE developed in the previous section along with the constraints of conformal symmetry.

Although they are primary fields, $E$ and $B$ need not be orthonormal fields.  In general, conformal invariance 
constrains the two-point function of two conserved currents to take the form
\begin{align}
\label{eq:JJ}
\expval{J^i_{(a)} (x) J^j_{(b)} (y)}&= \tau_{ab} \frac{I^{ij}(x-y)}{(x-y)^{4}}
\end{align}
where $I^{ij}(x)$ is the inversion tensor $\delta^{ij} - 2 x^i x^j / x^2$.  The two-point functions of our currents $E$ and $B$ are thus completely characterized by the three numbers $\tau_{EE}$, $\tau_{BB}$ and by Bose symmetry $\tau_{EB} = \tau_{BE}$.  
One consequence of parity is that if we require the theory be parity symmetric, then $\langle E^i(x) B^j (y) \rangle$ and its corresponding coefficient $\tau_{EB}$ must vanish, i.e.\ the currents are orthogonal.

To write down the bulk-boundary two-point functions
$\langle F_{\mu\nu}(x) E_i(y)\rangle$ and $\langle F_{\mu\nu}(x) B_j(y)\rangle$, we need some additional tensorial building blocks
compatible with conformal symmetry \cite{McAvity:1995zd}.
We will use the inversion tensor $I_{\mu\nu}$ defined below (\ref{eq:JJ}) along with its index contracted form $I_\mu \equiv I_{\mu\nu} z^\nu$ where $z^\mu$ is a boundary polarization vector satisfying $z \cdot n = 0$.  We need two additional structures as well\footnote{%
 The embedding space (or null cone) formalism can also be useful for writing down these correlation functions \cite{Billo:2016vm,Costa:2011wa,Costa:2014rya}.  We will
 use this formalism later in a development of the odd parity three-point functions but for the purposes here, a more 
 pedestrian approach suffices. 
}
\begin{align}
\begin{aligned}
X_{\mu}&=n_\mu -2\frac{x_\perp (x-y)_\mu}{(x-y)^2} \ ,
  & Y_{\nu}&=n_\nu +2\frac{y_\perp (x-y)_\nu}{(x-y)^2} \ .
\end{aligned}
\end{align}

The most general bulk-boundary correlation function between $F^{\mu\nu}$ and a boundary vector $J_{(a)}(y,z) = J^i_{(a)}(y)z_i$ takes 
the form
 \begin{align}
 	\expval{F_{\mu\nu}(x)J_{(a)}(y,z)}=\frac{1}{(2x_\perp)^{2-\Delta_J}(x-y)^{2\Delta_J}}\left(\alpha_{(a)} X_{[\mu}I_{\nu]}-i\beta_{(a)}\frac{\epsilon_{\mu\nu\rho\alpha}}{2}X^{\rho}I^{\alpha}\right) \ .
 \end{align}
The Maxwell equations for $F_{\mu\nu}$ impose that $\Delta_J=2$ (as does the condition that $J^\mu$ is a conserved current).
The Maxwell field is special in this respect; curiously, bulk-symmetric-tensor operators have a zero two-point function with boundary conserved currents (see e.g.\ \cite{Herzog:2021spv}).  

Now comparing the BOE of $F_{\mu\nu}$ (\ref{eq:Fope}), the two-point function of a conserved current (\ref{eq:JJ}), and the above expression, we see that
\begin{align}
\langle F_{\mu\nu}(x) E(y) \rangle &= \frac{1}{(x-y)^{4}} \left(\tau_{EE} X_{[\mu}I_{\nu]}-i \tau_{EB} \frac{\epsilon_{\mu\nu\rho\alpha}}{2}X^{\rho}I^{\alpha}\right) \ , \\ 
\langle F_{\mu\nu}(x) B(y) \rangle &=  \frac{1}{(x-y)^{4}} \left(\tau_{EB} X_{[\mu}I_{\nu]}-i \tau_{BB} \frac{\epsilon_{\mu\nu\rho\alpha}}{2}X^{\rho}I^{\alpha}\right) \ .
\end{align}
 
Next, we consider $\langle F_{\mu\nu} (x) F_{\lambda\rho} (y) \rangle$.  Bulk two-point functions can depend on an invariant
cross ratio
\begin{align}
	\xi = \frac{(x-y)^2}{4x_\perp \ y_\perp } \ .
\end{align}
The additional indices require more complicated tensor structures, still constructed from the previously defined
 building-blocks:\footnote{%
  One could ask whether the structures ${\mathcal R}^{(i)}_{\mu\nu, \rho \sigma} = \frac{i}{2} {\mathcal F}^{(i)}_{\rho \sigma, \alpha \beta}  {\epsilon^{\alpha \beta}}_{\rho \sigma}$ and $\widetilde {\mathcal F}^{(i)}_{\mu\nu, \rho \sigma} = 
  - \frac{1}{4}  {\mathcal F}^{(i)}_{\alpha \beta,\gamma \delta} {\epsilon^{\alpha \beta}}_{\mu\nu} {\epsilon^{\gamma \delta}}_{\rho \sigma}$ are independent of the ones listed in (\ref{eq:FFstructs}).  In fact, it is straightforward to check the linear dependencies
 \begin{align*}
{\mathcal L}^{(1)}_{\mu\nu,\rho\sigma} + {\mathcal R}^{(1)}_{\mu\nu,\rho\sigma} &= 0 \ , \; \; \; &
 {\mathcal L}^{(2)}_{\mu\nu,\rho\sigma} - {\mathcal R}^{(2)}_{\mu\nu,\rho\sigma} &= 2 {\mathcal L}^{(1)}_{\mu\nu,\rho\sigma} \ , \\
\widetilde {\mathcal F}^{(1)}_{\mu\nu,\rho\sigma} &= {\mathcal F}^{(1)}_{\mu\nu,\rho\sigma}  \ , \; \; \; &
 \widetilde {\mathcal F}^{(2)}_{\mu\nu,\rho\sigma} &= -{\mathcal F}^{(2)}_{\mu\nu,\rho \sigma} + 2{\mathcal F}^{(1)}_{\mu\nu,\rho\sigma} \ .
\end{align*}
 }
\bal
\label{eq:FFstructs}
	{\mathcal F}^{(1)}_{\mu\nu, \rho \sigma} &\equiv \frac{1}{2} (I_{\mu\rho} I_{\nu \sigma} - I_{\mu \sigma} I_{\nu \rho} )\ , \\
	{\mathcal F}^{(2)}_{\mu\nu, \rho \sigma} &\equiv  \frac{\xi}{1+\xi}\left(I_{\mu\rho} X_\nu Y_\sigma - I_{\mu\sigma} X_\nu Y_\rho - I_{\nu \rho} X_\mu Y_\sigma + I_{\nu \sigma} X_\mu Y_\rho \right)\ , \\
	{\mathcal L}^{(i)}_{\mu\nu, \rho\sigma} &\equiv \frac{i}{2} {\mathcal F}^{(i)}_{\alpha \beta, \rho\sigma} {\epsilon^{\alpha \beta}}_{\mu\nu}  \ .
\eal
It is now straightforward to write the general form of the bulk-to-bulk correlator,
\begin{equation}
\begin{aligned}
	\expval{F_{\mu\nu}(x)F_{\rho\sigma}(y)}=
	\frac{
	f_1(\xi)\mathcal{F}^{(1)}_{\mu\nu,\rho\sigma}+f_2(\xi)\mathcal{F}^{(2)}_{\mu\nu,\rho\sigma}
	+\widetilde{f_1}(\xi)\mathcal{L}^{(1)}_{\mu\nu,\rho\sigma} +\widetilde{f_2}(\xi)\mathcal{L}^{(2)}_{\mu\nu,\rho\sigma}
	}
	{(x-y)^4} \ .
\end{aligned}
\end{equation}
The functions $f_i$ and $\widetilde f_i$ are constrained by the conservation and Bianchi identities, with the generic solutions
\begin{align}
f_1(\xi) &= c_1 + c_2 \frac{\xi^2}{(1+\xi)^2} \ , & f_2(\xi) &= -c_2 \frac{\xi^2}{(1+\xi)^2} \  , \\
\widetilde f_1(\xi) &= \widetilde c_1 + \widetilde c_2 \frac{\xi^2}{(1+\xi)^2} \ ,& \widetilde f_2(\xi) &= -\widetilde c_2 \frac{\xi^2}{(1+\xi)^2}  \ .
\end{align}
That the solutions for $f_i$ and $\widetilde f_i$ are so similar follows from the fact that $F_{\mu\nu}$ and $\widetilde F_{\mu\nu}$ obey the same equations of motion.

We can fix the constants $c_i$ and $\widetilde c_i$ by matching to the leading terms in the BOE of the Maxwell field.  The result is
\begin{align}
\label{eq:crels}
c_1 = \tau_{EE} + \tau_{BB} \ , \; \; \; c_2 = \tau_{BB} - \tau_{EE} \ , \; \; \; \widetilde c_2 = - 2 \tau_{EB} \ , \; \; \; \widetilde c_1 = 0  \ .
\end{align}
In fact we have checked that the full
expression for $\langle F_{\mu\nu}(x) F_{\mu\nu}(y) \rangle$ can be recovered 
by using the BOE of $F_{\mu\nu}$ twice on the current-current correlation functions of $E$ and $B$ and resumming, but we omit the details.  

The equations of motion allow for a constant term $\widetilde{c}_1$ in $\widetilde f_1$, which is never realised. 
The vanishing of $\widetilde c_1$ can be explained as follows. Far away from the boundary, this correlator must reduce to pure free Maxwell. In this limit, only the structures $\mathcal{F}^{(1)}$ and $\mathcal{L}^{(1)}$ persist, and one must 
take the cross-ratio $\xi$ to zero. As the free two-point function without a boundary is parity-invariant 
(the $\theta$ term is a total derivative), $\widetilde{c}_1$ vanishes and only $c_1$ contributes. 

In fact, $c_1$ is fixed as well. In Feynman gauge, ignoring the boundary, we would normalize such that
\[
 \langle A_\mu(x) A_\nu(0) \rangle = \delta_{\mu\nu} \frac{g^2}{4\pi^2} |x|^{-2} \ ,
 \] 
 which fixes $c_1$ in terms of the gauge coupling.
 We find %
 \begin{align}
 \label{eq:c1g2rel}
 	c_1
	= \frac{4}{\pi {\rm Im}(\tau)}=\frac{2g^2}{\pi^2} 
	\ . 
 \end{align}
Via (\ref{eq:crels}), we find in turn a constraint on $\tau_{EE} + \tau_{BB}$ which we will discuss briefly
in section \ref{sec:moduli} and at length
in section \ref{sec:discussion}.\footnote{%
 The Fourier transformed form of the current correlation function is
 \[
 \tau_{ab} \frac{\pi^2}{2} \frac{ k_\mu k_\nu - \delta_{\mu\nu} k^2 }{k^2}  \ .
 \]
 Defining $ \sigma_{ab} \equiv \tau_{ab} \frac{\pi^2}{2} $, the sum rule is slightly nicer in Fourier transformed language,
 $\sigma_{EE} + \sigma_{BB} = g^2$. 
 \label{footnote:normalization}
}

  In what follows, we will swap the $c_i$, $\widetilde c_i$, and $\tau_{ij}$ for the new variables $\kappa$, $\chi_e$, and $\chi_o$ where
 \begin{align}
 \label{eq:kappachidefs}
 \kappa \equiv c_1 = \tau_{EE} + \tau_{BB} \ , \; \; \; \chi_e \equiv \frac{c_2}{c_1} = \frac{\tau_{BB} - \tau_{EE}}{\tau_{BB} + \tau_{EE}} \ , \; \; \; \chi_o \equiv \frac{\widetilde c_2}{c_1}=- \frac{2 \tau_{EB}}{\tau_{BB} + \tau_{EE}} \ .
 \end{align}
 With these new variables, we can write the $F_{\mu\nu}$ two-point function in the form 
 \begin{align}
	\expval{F_{\mu\nu} (x) F_{\rho\sigma}(y)}=\kappa \left( \frac{\mathcal{F}^{(1)}_{\mu\nu,\rho\sigma}}{(x-y)^{4}}+ \chi_e\frac{\mathcal{F}^{(1)}_{\mu\nu,\rho\sigma}-\mathcal{F}^{(2)}_{\mu\nu,\rho\sigma}}{(\widetilde{x}-y)^4}+\chi_o\frac{\mathcal{L}^{(1)}_{\mu\nu,\rho\sigma}-\mathcal{L}^{(2)}_{\mu\nu,\rho\sigma}}{(\widetilde{x}-y)^4}\right) \ ,
\end{align}
with $\tilde{x}^\mu \equiv x^\mu-2 x_\perp n^\mu$. The tensorial structures have an interpretation as an (even or odd) reflection of the initial bulk structure. The first term corresponds to the free two-point function, and so to the exchange of the identity in the bulk channel. 
The two remaining terms are due to the exchange of the bulk scalar and pseudo-scalar 
in the bulk OPE of $F_{\mu\nu}F_{\rho\lambda}$,
\begin{align}
\label{eq:FFonepoint}
	\expval{{:}F^{\mu\nu}F_{\mu\nu}(x){:}} &=6\kappa\frac{\chi_e}{(2x_\perp)^4}  \ , \; \; \;
	\expval{{:}F_{\mu\nu}\widetilde{F}^{\mu\nu}(x){:}} =6\kappa\frac{i \chi_o}{(2x_\perp)^4}  \ ,
\end{align}
which expectation values define a moduli space. We refer to $\chi_e$ and $\chi_o$ as reflection coefficients. 

The system of two-point functions of $E$ and $B$ must be reflection positive for a unitary CFT, which imposes the following two constraints on the $\tau_{ij}$:
\begin{align}
\tau_{EE} + \tau_{BB} \geq 0 \ , \; \; \; \tau_{EE} \tau_{BB} \geq \tau_{EB}^2 \ .
\end{align}
The constraint $\tau_{EE} + \tau_{BB} \geq 0$ we now see is equivalent to the positivity of $g^2$
via (\ref{eq:crels}) and (\ref{eq:c1g2rel}).  The second constraint, however, is less trivial and restricts the conformal manifold:
\begin{align}
\chi_e^2 + \chi_o^2 \leq 1 \ .
\end{align}
In other words, the moduli space of possible Maxwell field two-point functions can be parametrized by a disk of unit radius (corresponding to the possible values of $\chi_e$ and $\chi_o$) cross the half line (corresponding to the possible values of $g^2$ or equivalently $\kappa$).  

 In the limiting case $\chi_e^2 + \chi_o^2 = 1$, the matrix of current two-point functions becomes degenerate and $E$ and $B$ are the same up to rescaling, i.e.\ they are parallel.  We discussed in section \ref{sec:perturbativeperspective} that in free theories,
 some linear combination of the currents 
$E$ and $B$ must vanish.  Thus we conclude that all free theories must lie on the circle
 $\chi_e^2 + \chi_o^2 = 1$.  The converse is in fact also true, that if a theory lies on $\chi_e^2 + \chi_o^2 = 1$, then the 
 correlation functions of the remaining vector field must all follow from Wick's Theorem, as we will demonstrate later
 in appendix \ref{sec:perimeterimpliesfree}.

\subsection{Moduli Space \& Symmetries}
\label{sec:moduli}

The moduli space coordinatized by $(\chi_e, \chi_o, \kappa)$ describes
 a conformal manifold of theories. We now discuss how symmetries and dualities relate to features of the moduli space.

\begin{figure}[h!]
  \begin{center}
    \includegraphics[width=0.8\textwidth]{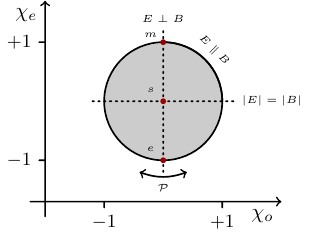}
  \end{center}
  \vspace*{-1.5em}
  \caption{Space of boundary conditions for the two-point function $\expval{F_{\mu\nu}(x) F_{\rho\lambda}(y) }$. The outer region is called the perimeter, and has $E\parallel B$. The center-point $s$ is mapped to itself under S-duality. The poles $m$ and $e$ correspond to purely magnetic and purely electric boundary conditions. The parity invariant meridian satisfies $E\perp B$. }
  \label{fig:phase2}
\end{figure}

Recall that under parity, $E$ is even, while $B$ is a pseudo-vector, hence $(\chi_e,\chi_o) \overset{\mathcal{P}}{\rightarrow}(\chi_e,-\chi_o)$.
In parity preserving theories, to which we will specialise later on, $\tau_{EB}$ and correspondingly $\chi_o$ must vanish, i.e.\ $E$ and $B$ are perpendicular. 
Theories with parity are restricted to the meridian line $-1 \leq \chi_e \leq 1$ and $\chi_o = 0$. 
At the point $\chi_e = 1$, $\tau_{EE}$ and the electric current must vanish, while at the point $\chi_e=-1$, $\tau_{BB}$ and the magnetic current must vanish.
Restoring $\chi_o$, we see that it encodes the strength of the parity breaking. 
In the maximally parity breaking cases ($\chi_o = \pm 1$ and $\chi_e = 0$), the currents 
$E$ and $B$ become the same up to a choice of sign. 
Along the perimeter $\chi_o^2 + \chi_e^2 = 1$, the $E$ and $B$
fields are parallel, which is realised clearly through the $\theta$ angle appearing in the proportionality constant between the currents in the case where the bulk decouples from the boundary degrees of freedom.   The story is summarized in fig.\ \ref{fig:phase2}.
%
%In the extreme case of the perimeter, this is realised clearly through the $\theta$ angle appearing in the proportionality constant between the currents. More generally, the theories with parity are restricted to the meridian line $-1 \leq \chi_e \leq 1$ and $\chi_o = 0$. 
%At the point $\chi_e = 1$, $\tau_{EE}$ and the electric current must vanish, while at the point $\chi_e=-1$, $\tau_{BB}$ and the magnetic current must vanish.
%Restoring $\chi_o$, we see that it encodes the strength of the parity breaking. 
%In the maximally parity breaking cases ($\chi_o = \pm 1$ and $\chi_e = 0$), the currents 
%$E$ and $B$ become the same up to a choice of sign.   The story is summarized in fig.\ \ref{fig:phase2}.
%

It is interesting to consider the action of SL$(2, {\mathbb Z})$ on this conformal manifold. 
We can deduce how SL$(2, {\mathbb Z})$ acts on $(\kappa, \chi_e, \chi_o)$ by looking at how
it acts on the field strength $\langle {:} F^2 {:} \rangle$  and $\langle {:} \tilde F^2 {:} \rangle$
one-point functions, which we now review.  
Let
\[
\left( \begin{array}{cc}
a & b \\
c & d 
\end{array}
\right)  \  \; \; \mbox{where} \; \; \; a, b, c, d \in {\mathbb Z}   \ \; \; \mbox{and}  \; \;   a b - cd = 1 
\]
be an element of SL$(2, {\mathbb Z})$ acting on the gauge coupling $\tau$ in the usual way
\[
\tau \to \frac{a \tau + b}{c \tau + d} \ .
\]
Introduce the self dual and anti-self dual combinations of the field strength:
$ F_{\mu\nu}^{\pm}=\frac{1}{2}\left(F_{\mu\nu}\pm i \widetilde F_{\mu\nu} \right)$.
From (\ref{eq:FFonepoint}), the one-point functions take the form
\begin{align}
	&\begin{aligned}
		\expval{{:}F_{+}^2(x){:}} &=3\frac{\kappa \chi}{(2x_\perp)^4}  \ , \; \; \;
	\expval{{:}F_{-}^2(x){:}} =3\frac{\kappa \overline{\chi}}{(2x_\perp)^4} 
		\end{aligned}
	 \ ,
\end{align}
defining a complexified reflection coefficient $\chi = \chi_e + i \chi_o$. The field-strengths $F^{\pm}$ 
have a simple transformation law, 
\begin{align}
    F^{+} &\to (c\overline{\tau}+d)F^{+}  \ , &  F^{-} &\to (c\tau+d)F^{-}  \ ,
\end{align}
which translates into an induced action on the moduli space:
\begin{align}
	\kappa &\to \kappa(c\tau +d)(c\overline{\tau}+d) \ , & \chi &\to \left(\frac{c\overline{\tau}+d}{c\tau+d}\right) \chi\ .
\end{align}
In other words, SL$(2, {\mathbb Z})$ acts as a phase rotation on the disk parametrized by $\chi$ and a rescaling of 
$\kappa$.  Under the action of S-duality, $\tau \to -1/\tau$,
the origin of the disk $\chi=0$ is a fixed point while $\kappa$ gets rescaled by the coupling $\kappa \to | \tau |^2 \kappa$. 
On the other hand, under $T: \tau \to \tau + b$, both $\kappa$ and $\chi$ are fixed. 
In the limit where we stay on the perimeter of the disk, this transformation is related to the SL$(2,\mathbb{Z})$ action on the $3d$ CFT of \cite{Witten:2003ya}, see \cite{DiPietro:2019hqe}.

 Before closing, it is important to emphasize the physical significance of these constraints $\tau_{EE} + \tau_{BB} = 2 g^2/ \pi^2$ and $\tau_{EE} \tau_{BB} \geq \tau_{EB}^2$.  The quantities $\tau_{EE}$ and $\tau_{BB}$ determine the charge conductivities of the matter degrees of freedom with respect to the two currents $E$ and $B$.  From these constraints, we deduce that in any theory of this type where the interactions are mediated by the bulk photon, both conductivities are positive numbers and that their magnitudes are bounded above by the coupling constant $2 g^2 / \pi^2$.  This sum rule was already noted in \cite{DiPietro:2019hqe} and a complexified version of it was rediscovered from localization in \cite{KumarGupta:2019nay}.  
 
We can push the reasoning a bit further and investigate what can happen for self-dual theories that map back to themselves under an element of SL$(2, {\mathbb R})$.  Suppose we are in a situation where the theory maps back to itself under a $TST$ transformation, where $T$ acts by a translation $\tau \to \tau + t$.  
 Two such theories were considered in \cite{KumarGupta:2019nay}, one with $t=0$ and the other with $t=1/2$
 (see also \cite{Bason:2023bin}).  
 Under such a transformation, $\tau = i \sqrt{1-t^2}$ is a fixed point for $0 \leq t < 1$, $\kappa$ is left unchanged,
 and $\chi$ rotates by a nontrivial phase given by $\cos^{-1} (1-2t^2)$.  Thus to be self-dual, we must fix $\chi=0$.  
 At the center of the disk, the $E$ and $B$ fields are on equal footing and have the same two-point functions,
 $\tau_{EE} = \tau_{BB}$.  Thus we conclude that at this proposed self-dual point, it must be that
 \be
 \tau_{EE} = \tau_{BB} = \frac{g^2}{\pi^2} = \frac{2}{\pi \sqrt{1-t^2}} \ . 
 \ee
 In other words, the constraint of self-duality is enough to completely fix the conductivities of these two currents.
A version of this result, formulated for scalar and fermionic matter degrees of freedom, can be found in \cite{Hsiao:2017lch, Hsiao:2018fsc}.

\section{A Bulk Constraint on Boundary Data}
\label{sec:threepoint}

As our ultimate goal is to bootstrap the correlation function of four currents, as a preliminary step we need to describe
the three-point functions of two currents with the operators that can be exchanged in the four-point correlator.
The exchanged operators are symmetric traceless tensors (STTs) $T_\ell$ where $\ell$ is the spin.  
Thus we begin this section by writing down the
$\langle J JT_\ell \rangle$ three-point functions compatible with conformal symmetry.

The currents in our problem are the boundary values of the same bulk field $F_{\mu\nu}$, 
and it turns out that regularity of the parent 
$\langle F_{\mu\nu} J T_{\ell} \rangle$ correlation function puts
nontrivial constraints on the boundary three-point functions.  In the second part of this section, we explain these constraints.

\subsection{The \texorpdfstring{$\langle J J T_\ell \rangle$}{<JJTl>} Three-Point Function}
\label{sec:JJT}

We first specify our basis choice for the different $\langle J J T_\ell \rangle$ correlation functions. We follow conventions close to \cite{He:2023ewx}.

\subsubsection*{Boundary Structures in \texorpdfstring{$\expval{J J T_\ell}$}{<JJTl>}}

We are interested in the configuration with two, potentially different, currents, and one STT operator of generic spin $\ell$. We will comment later on the specific cases $\ell =0,1$. We write the correlator in the index-free formalism \cite{Costa:2011wa}. We define
\begin{align}
	P_{mn}=-2P_{m}\cdot P_n = (x_m-x_n)^2  \ , \; \; \;
	Z_m \cdot P_n &= z_m \cdot x_{nm} \ ,
\end{align}
where we use $m$ and $n$ to index the insertion point.
 The even parity building blocks, in the embedding and real-space picture, take the form  
\begin{align}
	V_{l,mn}&=\frac{x_{lm}^2 z_l \cdot x_{ln}-x_{ln}^2 z_{l}\cdot x_{lm}}{\sqrt{x_{12}^2 x_{23}^2 x_{31}^2 }}= \frac{Z_{l}\cdot P_{m}P_{ln}-Z_{m}\cdot P_{n}P_{lm}}{\sqrt{P_{12}P_{23}P_{31}}}  \ , \\
	H_{mn} &= z_m\cdot z_n- 2 \frac{z_m \cdot x_{mn} z_n \cdot x_{mn}}{x_{mn}^2} = Z_m \cdot Z_{n}+2\frac{Z_m\cdot P_n Z_n \cdot P_m}{P_{mn}} \ ,
\end{align}
and all the structures in both frames indeed correspond exactly to one another under projection. We pick the shorthand convention 
\begin{align}
	V_{1}&=V_{1,23} \ , & V_{2}&=V_{2,31} \ , & V_{3}&=V_{3,12} \ .
\end{align}
The parity-odd structures are  more involved. In the embedding space they are 
\begin{align}
	\epsilon_{mn}=2i\frac{\epsilon(Z_m,Z_n,P_1,P_2,P_3)}{\sqrt{P_{12}P_{23}P_{31}}} \ .
\end{align}
It turns out these three building blocks are enough to generate all the parity-odd structures in the three-point functions.
In fact for $\ell \geq 2$, we can trade $\epsilon_{12}$ for the other two building blocks.

\subsubsection*{OPE Coefficients and Conventions for \texorpdfstring{$\expval{JJT_\ell}$}{<JJT>}}

In this subsection we specify our basis choice for the enumeration of the different structures. 
%Our conventions follow closely \cite{He:2023ewx}. 
Our setup is that of two conserved currents $E$ and $B$ which we collectively refer to as $J_{(a)}$ and a STT $T_\ell$:
\begin{align}
	\expval{J_{(a)}(x_1,z_1)J_{(b)}(x_2,z_2)T_\ell(x_3,z_3)}=\frac{\sum_{I} \gamma_{ab}{}^I E^{(I)} + \sum_{I} \widetilde \gamma_{ab}{}^{I} O^{(I)}}{x_{12}^{4-\Delta} x_{13}^{\Delta} x_{23}^{\Delta}} \ .
\end{align}
We now define the different structures $E^{(I)}$ and $O^{(I)}$.  We also specify the consequences of conservation and Bose symmetry on the OPE coefficients $\gamma_{ab}^I$ and $\widetilde \gamma_{ab}^I$.

\paragraph{Primary with \texorpdfstring{$\ell \geq 2$}{l>=2} :}

Given the $V_i$, $H_{ij}$ and $\epsilon_{ij}$ building blocks above, we can parametrise the three-point function as 
\begin{align}
\begin{aligned}
	E^{(1)}&= V_1 V_2 V_3^\ell  \ , & E^{(2)}&= V_1 H_{23}V_3^{\ell-1}  \ , \\
	E^{(3)}&= H_{13} V_2 V_3^{\ell-1} \ ,  &  E^{(4)}&= H_{13} H_{23} V_3^{\ell-2}  \ , \\
	E^{(5)}&= H_{12} V_3^\ell  \ ,
\end{aligned}
\end{align}
and
\begin{align}
\begin{aligned}
	O^{(1)}&= \epsilon_{13} V_2 V_3^{\ell-1} \ ,  & O^{(2)}&= \epsilon_{23} V_1 V_3^{\ell-1}  \ , \\
	O^{(3)}&= \epsilon_{13} H_{23} V_3^{\ell-2}  \ ,  &  O^{(4)}&=\epsilon_{23} H_{13} V_3^{\ell-2}  \ .
\end{aligned}
\end{align}
Note that we got rid of the $\epsilon_{12}$ structure, as it is redundant for $\ell \geq 2$, as follows from the identity  
\begin{align}
	V_3^2 \epsilon_{12} = \left(H_{13}-V_1 V_3 \right)\epsilon_{23}-\left(H_{23}-V_2 V_3 \right)\epsilon_{13} \ .
\end{align}

Conservation imposes the  constraints 
\bal
	\gamma^{5}&=\frac{(\Delta-2)\gamma^{1}+\left(4+\ell-\Delta\right)\gamma^2}{\Delta+\ell} \ , \\  
	\gamma^{4}&=\frac{\ell(\Delta-2)\gamma^{1}+\left(\ell(5+\ell)-\Delta\left(\Delta+2\ell-1\right)\right)\gamma^2}{(\Delta+\ell)(2+\ell-\Delta)}  \ , \\
	 \gamma^3 &= \gamma^2 \ .
\eal
and 
\begin{align}
	\widetilde{\gamma}^{4}&=\frac{\widetilde{\gamma}^1+(\Delta-2)\widetilde{\gamma}^2}{\Delta-\ell-3}  \ , & \widetilde{\gamma}^{3}&=\frac{(\Delta-2)\widetilde{\gamma}^1+\widetilde{\gamma}^2}{\Delta-\ell-3} \ .
\end{align}

Using Bose symmetry, one can swap the two currents to relate the matrix of coefficients:
\begin{align}
	\begin{pmatrix}
		\gamma_{ab}^1 \\
		\gamma_{ab}^2 
	\end{pmatrix} &= (-1)^{\ell}\begin{pmatrix}
		1 & 0 \\ 0 & 1 
	\end{pmatrix}\begin{pmatrix}
		\gamma_{ba}^1 \\
		\gamma_{ba}^2 
	\end{pmatrix}   \ , & \begin{pmatrix}
		\widetilde{\gamma}_{ab}^1 \\
		\widetilde{\gamma}_{ab}^2 
	\end{pmatrix} &= (-1)^{\ell+1}\begin{pmatrix}
		0 & 1 \\ 1 & 0
	\end{pmatrix}\begin{pmatrix}
		\widetilde{\gamma}_{ba}^1 \\
		\widetilde{\gamma}_{ba}^2 
	\end{pmatrix} \ .
\end{align}
This crossing equation reduces the amount of data we have to consider. To wit, if we consider $\expval{EET_\ell}$, we find that $\ell = 2n$ implies $\widetilde{\gamma}_{EE}^2=-\widetilde{\gamma}_{EE}^{1}$; $\ell=2n+1$ implies $\gamma_{EE}^{I}=0$ and $\widetilde{\gamma}_{EE}^2=+\widetilde{\gamma}_{EE}^{1}$. The constraints for $\expval{BBT_\ell}$ are the same. 
These results will be important for the bootstrap application later because they mean that
for each even spin $\ell \geq 2$, there are three independent OPE coefficients, which we can take to be
$\gamma_{EE}^1$, $\gamma_{EE}^2$ and $\widetilde \gamma_{EE}^1$.  However for odd spin $\ell \geq 3$,
there is just $\widetilde \gamma_{EE}^1$.

In view of the bootstrap application to come, we introduce $\gamma^+_{EE}$ and $\gamma^-_{EE}$ where
\bal
\gamma^1_{EE} &= 2 (1 + \Delta) (\Delta + 2 \Delta \ell^2 + \ell (5 + 2 \ell (1 + \ell)))\gamma_{EE}^- + ( 
    \Delta - \ell-2) (3 \Delta + \ell-4) \gamma_{EE}^+ 
 \ , \\
\gamma^2_{EE} &= 2 (1 + \Delta) \ell (2 + \Delta (\ell-1) + \ell^2) \gamma_{EE}^- + ( \Delta - \ell-2) ( 
    2 \Delta + \ell-2) \gamma_{EE}^+ \ .
\eal 
Note one needs to be a bit careful with these relations.  While the map from $\gamma^\pm$ to $\gamma^I$
is generically full rank, it can be that for certain specific values of $\Delta$, the rank decreases from two to one, in which
case SDPB does not function well.\footnote{%
 Personal communication Petr Kravchuk. 
}
We have designed the transformations such that the rank is always two for $\Delta$ in the unitary region.  

For a mixed current bootstrap, we will ultimately 
need the constraints $\gamma_{EB}^{I}=(-1)^{\ell}\gamma_{BE}^{I}$, and $(\widetilde{\gamma}_{EB}^1\pm \widetilde{\gamma}_{EB}^2)=\pm (-1)^{\ell+1}(\widetilde{\gamma}_{BE}^1 \pm \widetilde{\gamma}_{BE}^2)$,
but not in this work.

\paragraph{Primary with $\ell =1$ :}

We now note the modification when $\ell =1$. We have fewer structures in both sectors, but lose the linear relation that got rid of $\epsilon_{12}$. In this setting we use the basis
\bal
	E^{(1)}&= V_1 V_2 V_3  \ ,& E^{(2)}&= V_1 H_{23}  \ ,\\
	E^{(3)}&= H_{13} V_2   \ , &  	E^{(4)}&= H_{12} V_3  \ ,
\eal
and 
\bal
	O^{(1)}&= \epsilon_{13} V_2  \ , & O^{(2)}&= \epsilon_{23} V_1  \ ,  & O^{(3)}&= \epsilon_{12}  V_3 \ .
\eal
which defines what we mean by $\gamma^I$ and $\widetilde \gamma^I$ when $\ell=1$. 

Imposing conservation imposes the relations
\bal
	\gamma^{4}&=\frac{\Delta-1}{\Delta+3}\gamma^1 \ , & \gamma^{3}&=\gamma^{2}=\frac{\gamma^{1}}{\Delta+3}  \ , & \widetilde{\gamma}^{3}&=(\Delta-3)\widetilde{\gamma}^{1}  \ , & \widetilde{\gamma}^2 &= -\widetilde{\gamma}^{1} \ .
\eal
Bose symmetry in turn implies the properties
\begin{align}
	\gamma_{ab}^{1} &= -\gamma_{ba}^{1} \ , & \widetilde{\gamma}_{ab}^{1} &= -\widetilde{\gamma}_{ba}^{1} \ .
\end{align}
Now, the resulting simplification is even more straightforward, as $\gamma_{EE}^{I}=0=\widetilde{\gamma}_{EE}^{I}$, and $\gamma_{EB}^{I}=-\gamma_{BE}^{I}$, $\widetilde{\gamma}_{EB}^{I}=-\widetilde{\gamma}_{BE}^{I}$, a degenerate case of the $\ell \geq 2$ conditions.
In other words, in the bootstrap application to come of four identical currents, 
we will not need to consider spin one exchange.
% previous installment.

\paragraph{Primary with $\ell =0$ :}
 In this simplest of the cases, we have the independent
structures
\begin{align}
	E^{(1)}&= V_1 V_2 \ , & E^{(2)}&= H_{12} \ , & O^{(1)}&= \epsilon_{12} \ .
\end{align}
Conservation again brings restriction on the data, 
\begin{align}
	\gamma^{2} &= \frac{\Delta-2}{\Delta}\gamma^{1} \ .
\end{align}
Bose symmetry implies
\begin{align}
	\gamma_{ab}^{1}&=\gamma_{ba}^{1} \ , & \widetilde{\gamma}_{ab}^{1}&=\widetilde{\gamma}_{ba}^{1}  \ ,
\end{align}
which is weaker than the previous case, yielding only  $\gamma_{EB}^{I}=\gamma_{BE}^{I}$ and $\widetilde{\gamma}_{EB}^{I}=\widetilde{\gamma}_{BE}^{I}$.
For the bootstrap, we will have one OPE coefficient $\gamma_{EE}^+ \equiv \frac{1}{\Delta} \gamma_{EE}^1$ for even parity scalar exchange and one OPE coefficient $\widetilde \gamma_{EE}^1$ for odd parity scalar exchange.

\subsection{The \texorpdfstring{$\langle F_{\mu\nu} J T_\ell \rangle$}{<FJTl>} Three-Point Function}

In preparation for finding constraints on $\langle J J T_\ell \rangle$ from  $\langle F_{\mu\nu} J T_\ell \rangle$, we
first need to find a way of representing this bulk-boundary-boundary correlator. 
It general, it can depend on a function of a cross ratio
\begin{align}
	v= \frac{(x_1 \cdot n)^2 x_{23}^2}{x_{13}^2 x_{12}^2} \ ,
\end{align}
which goes to zero on the boundary. There are six structures that we use to represent the
six tensorial components of $F_{\mu\nu}$ in this three-point function.
%, there are $6$ (even, and $6$ odd) antisymmetric structures through which we can parametrise
% the correlation function. 
To build the tensor structures, we first consider the bulk-to-boundary building blocks 
\begin{align}
	X_{2}^{\mu}&= n^\mu - 2\frac{x_{12}^{\mu} x_1 \cdot n}{x_{12}^2}  \ , & X_{3}^{\mu}&=n^\mu - 2\frac{x_{13}^{\mu} x_1 \cdot n}{x_{13}^2}  \ ,
\end{align}
as well as the analogue of $H_{ij}$ with one index freed,
\begin{align}
	H_{2}^{\mu}&= z_2^\mu - 2\frac{x_{12}^{\mu}z_2\cdot x_{12}}{x_{12}^2} \ , & H_{3}^{\mu}&= z_3^\mu - 2\frac{x_{13}^{\mu}z_3\cdot x_{13}}{x_{13}^2} \ .
\end{align}

From these building blocks are constructed the six allowed even parity objects
\begin{align}
	&X_{2}^{[\mu}X_{3}^{\nu]} \ , & &X_{2}^{[\mu}H_{2}^{\nu]} \ , & &X_{3}^{[\mu}H_{2}^{\nu]} \ , \\
	&X_{2}^{[\mu}H_{3}^{\nu]} \ , & &X_{3}^{[\mu}H_{3}^{\nu]} \ , & &H_{2}^{[\mu}H_{3}^{\nu]} \ ,
\end{align}
which we antisymmetrise with $2X^{[\mu}Y^{\nu]}=X^{\mu}Y^{\nu}-Y^{\mu}X^{\nu}$. 
One has their odd counterpart as well, which we define by acting with $\frac{i}{2}\epsilon_{\mu\nu\alpha\beta}$.

We can now define the correlator of interest
\begin{align}
	\expval{F_{\mu\nu}(x_1)J_{b}(x_2,z_2)T(x_3,z_3)}=\frac{\sum_{I} f_{b,I}(v)\mathcal{S}_{\mu\nu}^{(I)} + \sum_{I} g_{b,I}(v)\mathcal{K}_{\mu\nu}^{(I)}}{x_{12}^{4-\Delta} x_{13}^{\Delta} x_{23}^{\Delta}}  \  , 
\end{align}
where the functions $f_{b,I}$ and $g_{b,I}$ are called form factors and will be our main interest in what follows. For simplicity, we focus on the parity-even sector $f_{b,I}$.  The constraints on the parity even and parity odd sector are essentially the same because the Bianchi identity and equation of motion for $F_{\mu\nu}$ are exchanged under S-duality.

In total, there are nine different parity even objects that can enter in this correlator. They are given by 
\begin{equation}
\begin{aligned}
	\mathcal{S}^{(1)}_{\mu\nu}&=X_{2}^{[\mu}X_{3}^{\nu]}V_2 V_3^{\ell}  \ , & \mathcal{S}^{(2)}_{\mu\nu}&=X_{2}^{[\mu}X_{3}^{\nu]}H_{23} V_3^{\ell-1} \ ,  & \mathcal{S}^{(3)}_{\mu\nu}&=X_{2}^{[\mu}H_{2}^{\nu]} V_3^{\ell}\ ,  \\
	\mathcal{S}^{(4)}_{\mu\nu}&=X_{3}^{[\mu}H_{2}^{\nu]} V_3^{\ell} \ , & \mathcal{S}^{(5)}_{\mu\nu}&=X_{2}^{[\mu}H_{3}^{\nu]} V_2 V_3^{\ell-1} \ , & \mathcal{S}^{(6)}_{\mu\nu}&=X_{2}^{[\mu}H_{3}^{\nu]} H_{23} V_3^{\ell-2} \ , \\
	\mathcal{S}^{(7)}_{\mu\nu}&=X_{3}^{[\mu}H_{3}^{\nu]} V_2 V_3^{\ell-1} \ , & \mathcal{S}^{(8)}_{\mu\nu}&=X_{3}^{[\mu}H_{3}^{\nu]} H_{23} V_3^{\ell-2} \ , & \mathcal{S}^{(9)}_{\mu\nu}&=H_{2}^{[\mu}H_{3}^{\nu]}V_3^{\ell-1} \ .
\end{aligned}
\end{equation}
The nine parity odd objects $\mathcal{K}_{\mu\nu}^{(I)}=\frac{i}{2}\epsilon_{\mu\nu\alpha\beta}\mathcal{S}^{(I)\alpha\beta}$ follow from contraction with the epsilon-tensor.
Note that if $\ell=1$, one must set $f_{b,6}(v)=0$ and $f_{b,8}(v)=0$, and likewise for the odd components. If $\ell=0$, we set $f_{b,2}$, $f_{b,5}$,  $f_{b,7}$, and $f_{b,9}$ and their odd counterparts to zero as well.

%**********
%
%Note that the parity even component of the $E$ correlator contribute to this parity even sector. The parity odd sector of the $B$ correlator contribute also. Since the duality transformation between $F$ and $\widetilde{F}$ preserves all the equations of motions and the kinematics, and that this transformation is involutive, it must be that all the parity odd structures of the bulk-boundary-boundary correlator can be obtained through the contraction of the parity even structures by $\frac{1}{2}\epsilon_{\mu \nu \alpha \beta}$, which we will call $\mathcal{K}_{\mu\nu}^{(i)}$. It also follows, as in the $\expval{FF}$ case, that the functions multiplying a given structure $\mathcal{S}^{(i)}$ and its odd counterpart $\mathcal{K}^{(i)}$ must have the same functional form. 
%
%We note that if $\ell=1$, one must set $f_{b,6}(v)=0$ and $f_{b,8}(v)=0$, and likewise for the odd components. If $\ell=0$, we set $f_{b,2/5/6/7/8/9}(v)=0$.

The form factors $f_{a,I}$ and $g_{b,J}$ 
can be decomposed into a sum of boundary conformal blocks which are functions of the cross ratio $v$.
If the field theory is moved to hyperbolic space through a Weyl transformation then when $v=1$, the bulk insertion lies on a geodesic joining the two boundary insertions.  In general, this point $v=1$ marks the end of the region of convergence of the sum of conformal blocks.  However, in our case, the sum over conformal blocks is finite.  Only the two currents $E$ and $B$ are in the BOE of the Maxwell field.   There is no possibility
to eliminate any naive divergences at $v=1$ through an infinite sum.  Instead, divergences at $v=1$ have to be
cured by imposing additional constraints on the boundary data of the CFT.  It is these constraints we now aim to extract. Similar behavior was exploited for a free in the bulk scalar in the series of publications \cite{Lauria:2021ut,Behan:2020nsf,Behan:2021tcn}. This feature is generic, and gives rises to sum rules which have recently been studied in AdS for scalars and stress-tensor insertions \cite{Levine:2023ywq,Meineri:2023mps}.

\subsection{Resummation and OPE Constraint}
\label{sec:OPEconstraints}

We followed two separate pathways to obtain the bulk regularity constraints on the boundary OPE coefficients.
Both led to the same results.  
The first pathway was to apply Maxwell's equations and conservation 
to the $\langle F_{\mu\nu} J_{(a)} T_{\ell} \rangle$ three-point
function and solve the resulting coupled system of differential equations for the form factors $f_{a,I}$ and $g_{b,J}$.  
The second was to apply the BOE (\ref{eq:Fope}) to the boundary three-point function $\langle J_{(a)} J_{(b)} T_\ell \rangle$
to reconstitute  $\langle F_{\mu\nu} J _{(a)}T_{\ell} \rangle$.  The first approach requires some cleverness to reduce the system of differential equations to a smaller manageable subset, and still 
leaves a number of integration constants
that can only be fixed by referring to the BOE.  However, looking at just the first few terms of the BOE is sufficient
to fix all the undetermined quantitites.  The second approach leads to the answer directly but requires some cleverness 
to handle the infinite sums.  We will describe only the second approach below.

\subsubsection*{General Approach for Resummation}

To express the form factors using the boundary data, we need to act with the previously defined 
BOE differential operators (\ref{eq:DDidefs}) on the boundary three-point function. This highly technical endeavour we automated with Mathematica. Intermediate stages of the computation involve sums of thousands of hypergeometric functions along with tensorial structures. In this subsection, we sketch the procedure and give the resulting constraint obtained. 

Our starting point is the following kinematical configuration%
\begin{align}
\label{JJTlimit}
	\lim_{|x_3| \to \infty} |x_3|^{2 \Delta} \expval{J(x,z_1)J(0,z_2)T(x_3,z_3)}=\frac{\sum_{I} \gamma^I E^{(I)} + \sum_{I} \widetilde \gamma^I O^{(I)}}{\abs{x}^{4-\Delta}} \ ,
\end{align}
leaving the first point unfixed so that it can be acted upon by the differential operators in the BOE. 
In this setting, the different building blocks previously defined take the simplified form
\bal
	V_1 &\to -\hat{x}\cdot z_1  \ , & H_{12}&\to z_1 \cdot z_2 - 2 z_1 \cdot \hat{x}z_2\cdot \hat{x}  \ , & \epsilon_{12}&\to i\epsilon\left(z_1,z_{2},\hat{x}\right)  \ ,\\
	V_2 &\to -\hat{x}\cdot z_2   \ , & H_{13} &\to z_1\cdot \hat{z}_3   \ , &  \epsilon_{13}&\to i\epsilon\left(z_1,z_3,\hat{x}\right)   \ , \\
	V_3 &\to +\hat{x}\cdot \hat{z}_3  \ , &  H_{23} &\to z_2\cdot \hat{z}_3  \ , & \epsilon_{23}&\to i\epsilon\left(z_2,z_3,\hat{x}\right) \ ,
\eal
with $\hat{z}_3 = z_3 -2(\theta\cdot z_3)\theta$ entering in all these correlators, $\theta$ encodes the direction at infinity along which we send the third point. Since this vector satisfies $\hat{z}_3^2 = 0$ (because $\theta^2=1$), it is a bona-fide polarisation vector for the third point, and we will drop the hat on it from now on. We also defined $\hat{x}^{i}=\frac{x^{i}}{\sqrt{x\cdot x}}$.  One can play the same game with the bulk structures to match the final results to the form factors.

This configuration makes it straightforward to reproduce most of the OPE constraints we already stated. The swapping property is implemented by sending 
\[
(\hat{x}, z_1, z_2) \to (-\hat{x},z_2, z_1) \ .
\] 
Imposing conservation on the first insertion can be done by taking the divergence, and on the second one by performing the swap and then repeating the same operation.  The new OPE constraints we want to derive however, require more work.

Our goal is now to act on this correlator (\ref{JJTlimit}) with the boundary OPE of \eqref{eq:Fope}. 
We break the problem apart into smaller pieces and initially determine how the $k^{\rm th}$ power of the Laplacian acts on these building blocks.
The numerators in our tensor structures take the form
\begin{align*}
	i)&\quad  (\hat{x}\cdot z_3)^{m} & 
	ii)&\quad   \hat{x}\cdot q(\hat{x}\cdot z_3)^{m}  & 
	iii)& \quad  \hat{x}\cdot q \  \hat{x}\cdot p(\hat{x}\cdot z_3)^{m}
\end{align*}
for some number $m$ and constant vectors $p$ and $q$. In these expressions and the following ones, $x$ stands for the tangential components of the position, and so does $x\cdot x$ and $x^2$, except when appearing as $x_\perp$. Consider for example
\begin{align}
	\mathcal{D}(x_\perp,\partial) \frac{(x\cdot z_3)^{m}}{(x\cdot x)^{2+\frac{m-\Delta}{2}}} \ , 
\end{align}
which can be computed using the identity
\begin{align}
	\Box^{k}  \frac{(x\cdot z_3)^{m}}{(x\cdot x)^{2+\frac{m-\Delta}{2}}} &= 4^k\left(\frac{3-m-\Delta}{2}\right)_{k}\left(\frac{4+m-\Delta}{2}\right)_{k}\frac{(x\cdot z_3)^{m}}{(x\cdot x)^{2+\frac{m-\Delta}{2}+k}} \ .
\end{align}
The $\Box^k$ operator has returned an expression of the same general form but with shifted exponent in the denominator. The remaining resummation is of hypergeometric type, and similar to other conformal block computations, see e.g. \cite{Dolan_2001,Dolan:2004up,Dolan:2012wt,McAvity:1995zd,Billo:2016vm,Herzog:2022jlx,Behan:2020nsf,Behan:2021tcn}. This gives
\begin{align}
\label{eq:Dresult}
	\mathcal{D} \frac{(x\cdot z_3)^{m}}{(x\cdot x)^{2+\frac{m-\Delta}{2}}} &= \frac{(\hat{x}\cdot z_3)^{m}}{(x^2+x_\perp^2)^{2-\Delta/2}}\frac{\, _2F_1\left(\frac{3-m-\Delta}{2},\frac{4+m-\Delta}{2};\frac{1}{2};\frac{v}{v-1}\right)}{(1-v)^{2-\Delta/2}} \ .
\end{align}
There is a similar result from ${\mathcal D}^i$ acting on $(x \cdot z_3)^m / |x|^{2 \alpha}$:
\begin{equation}
\begin{aligned}
\label{eq:Diresult}
	\mathcal{D}^{i}  \frac{(x\cdot z_3)^{m}}{(x\cdot x)^{2+\frac{m-\Delta}{2}}} &=\frac{(\hat{x}\cdot z_3)^{m-1}}{(x^2+x_\perp^2)^{2-\Delta/2}}\frac{\sqrt{v}}{\sqrt{1-v}^{\Delta-5}}\bigg(m z_3^{i} \, _2F_1\left(\frac{3-m-\Delta}{2} ,\frac{4+m-\Delta}{2};\frac{3}{2};\frac{v}{v-1}\right)\\
	&+\hat{x}^{i}(\Delta-m-4) z_3\cdot \hat{x}\, _2F_1\left(\frac{3-m-\Delta}{2},\frac{6+m-\Delta}{2};\frac{3}{2};\frac{v}{v-1}\right)\bigg) \ .
\end{aligned}
\end{equation}
To replace the numerator $(x \cdot z_3)^m$ with its more complicated cousins $\hat{x}\cdot q\, (\hat{x}\cdot z_3)^{m}$
and $\hat{x}\cdot q \  \hat{x}\cdot p\, (\hat{x}\cdot z_3)^{m}$, we use the identities
\begin{align}
\label{eq:qxzx}
	\frac{q\cdot \hat{x} (\hat{x}\cdot z_3)^{m}}{(x\cdot x)^{2-\frac{\Delta}{2}}}&= \frac{q\cdot\partial}{\Delta-m-3}\left(\frac{(\hat{x}\cdot z_3)^{m}}{(x\cdot x)^{2-\frac{(\Delta+1)}{2}}}\right)+\frac{m (q\cdot z)}{3+m-\Delta}\left( \frac{(\hat{x}\cdot z_3)^{m-1}}{(x\cdot x)^{2-\frac{\Delta}{2}}}\right)  \ , \\
\label{eq:pxqxzx}
	\frac{p\cdot \hat{x} \ q\cdot \hat{x} (\hat{x}\cdot z_3)^{m}}{(x\cdot x)^{2-\frac{\Delta}{2}}}&= \frac{p\cdot\partial}{\Delta-m-4}\left(\frac{q\cdot \hat{x}(\hat{x}\cdot z_3)^{m}}{(x\cdot x)^{2-\frac{(\Delta+1)}{2}}}\right)+\frac{m (p\cdot z)}{4+n-\Delta}\left( \frac{q\cdot \hat{x}(\hat{x}\cdot z_3)^{m-1}}{(x\cdot x)^{2-\frac{\Delta}{2}}}\right) \nonumber\\
	 &+\frac{(p\cdot q)}{4m-\Delta}\left( \frac{(\hat{x}\cdot z_3)^{m}}{(x\cdot x)^{2-\frac{\Delta}{2}}}\right) \ .
\end{align}

Indeed, the result of acting on these rational expressions with ${\mathcal D}$ and ${\mathcal D}_i$ 
will be linear combinations of  hypergeometric function multiplied by power laws. 
Looking already at (\ref{eq:Dresult}), 
there is clearly a potential divergence at $v=1$, in this case both from the power law prefactor and the ${}_2 F_1$ itself.  
The extra constraints on the OPE coefficients $\gamma^I$ and $\widetilde \gamma^I$ are needed to cure these potential divergences.

With the pieces (\ref{eq:Dresult}), (\ref{eq:Diresult}), (\ref{eq:qxzx}), and (\ref{eq:pxqxzx}) in hand, it becomes an algorithmic although tedious task to bring together these different elements, perform the resummation, and identify the different form factors. As they are sums of multiple hypergeometric functions with shifted arguments, their explicit expressions are not particularly enlightening. We performed this summation using Mathematica, and will now only quote the result for the constraint coming from bulk regularity. Note that in all of these different cases, there are multiple non-trivial checks, as there are more functions than variables to be constrained, and their Taylor series around the problematic $v=1$ points admit multiple poles, which all must cancel after fixing $1$ or $2$ linear relations. Moreover, there are no approximations made at any steps of our procedure; this is a purely analytical computation.

\subsubsection*{Results for \texorpdfstring{$\ell = 0$}{l=0}}

The resummation yields three independent form factors that depend on two parameters, $\gamma_{ab}^{1}$ and $\widetilde{\gamma}_{ab}^{1}$. A single linear relation is enough to make all even functions regular, 

\begin{align}
	\widetilde{\gamma}_{Bk}^{1}&=\underbrace{\frac{(\Delta-1)(\Delta-3)}{(\Delta-2)\Delta}\tan(\frac{\pi\Delta}{2})}_{f(\Delta)}\gamma_{Ek}^{1} \ ,
\end{align}
while the constraint coming from the odd form factors can be obtained by using an Hodge-transformation to swap $E \to B$, $B \to -E$, giving the dual relation 
\begin{align}
	\widetilde{\gamma}_{Ek}^{1}&=-f(\Delta)\gamma_{Bk}^{1} \ .
\end{align}

We stress that we do not ask Hodge-duality to be a symmetry of our theory; we merely use it as a convenient tool to extract the consequences of the odd sector from the ones  of the even sector. This relation coupled with Bose symmetry allows us to express all the OPE coefficients involving a $B$ field in terms of the OPE coefficients involving only the $E$ field:
\bal
\label{eq:gammascalar}
	\gamma_{BB} &= - \gamma_{EE}  \ , & \gamma_{BE} &= \frac{1}{f(\Delta)} \widetilde{\gamma}_{EE}\ , \\
	\widetilde{\gamma}_{BB} &= - \widetilde{\gamma}_{EE}  \ , & \widetilde{\gamma}_{BE} &= f(\Delta)\gamma_{EE} \ .
\eal
The net effect is that the bootstrap of $\expval{EEEE}$ is left unchanged by the existence of the bulk. However the data of $\expval{EEEB}$, $\expval{EEBB}$, $\expval{EBBB}$, and $\expval{BBBB}$ are non-trivially related to $\expval{EEEE}$, which may generate further bounds. It is non-trivial that the data of $\expval{EEEE}$ can, provided these identifications, also satisfy crossing symmetry for these other correlators.
As a quick example, we see that the bootstrap of $\expval{BBBB}$ is equivalent to the one of $\expval{EEEE}$, since the OPE coefficients appear squared.

For specific values of $\Delta$, $f(\Delta)$ may vanish or diverge, in which case one of the OPE coefficients in the relation must vanish.  
These special cases occur for $k$ a non-negative integer and
%All of these relations are valid provided $f(\Delta)$ is non-zero and finite. These relations will fail 
%for $k$ a non-negative integer if 
%
\begin{align}
	f(\Delta)=0 \; \, \leftrightarrow \Delta = 4+2k \ , \\
	f(\Delta)=\infty  \leftrightarrow \Delta = 5+2k \ .
\end{align}
In the cases $\Delta$ is even, the odd parity $\widetilde \gamma$ coefficients vanish.
In the cases $\Delta$ is odd, the even parity $\gamma$ coefficients vanish.  In either of these
cases, we lose the glue that relates $\gamma_{EE}$ to $ \gamma_{BB}$ (correspondingly $\widetilde \gamma_{EE}$ to $\widetilde \gamma_{BB}$); 
we cannot necessarily conclude that for the remaining nonzero OPE coefficients $\gamma_{EE} = -\gamma_{BB}$ or $\widetilde \gamma_{EE} = - \widetilde \gamma_{BB}$.  

\subsubsection*{Results for \texorpdfstring{$\ell = 1$}{l=1}}

There are seven form factors with potential divergences $v=1$.  
A single linear relation is enough to get rid of the poles:
%Each form factor has poles of multiple order (roughly $4$), which all must cancel at $v=1$ 
%by imposing a single constraint. It turns out that the relation 
%
\begin{align}
	\widetilde{\gamma}_{Bk}^{1}&=\underbrace{\frac{(\Delta -2) \Delta  }{(\Delta -3)(\Delta -1)(\Delta +3)}\tan(\frac{\pi (\Delta +1)}{2})}_{f_1(\Delta)}\gamma_{Ek}^{1} \ .
\end{align}
Bose symmetry puts tight constraints on the OPE coefficients in the $\ell=1$ case, which in combination with the regularity constraint now impose
%Using the previously stated behaviour under swapping, 
%we now find that the relation between the magnetic and electric sector follows through 
%
\begin{align}
	0=\gamma_{EE} = \frac{1}{f_1(\Delta)}\widetilde{\gamma}_{BE} \ ,  \\
	0 = \widetilde{\gamma}_{BB} = f_1(\Delta) \gamma_{EB}  \ .
\end{align}
As a result, the mixed coefficients $\gamma_{EB}$ and $\widetilde \gamma_{EB}$ can be nonzero
only in the special cases where $f_1(\Delta)$ vanishes or diverges, 
\begin{align}
	 \gamma_{EB} \neq 0 \Rightarrow f_1(\Delta)&=0 \; \, \leftrightarrow \Delta=5+2k \ ,  \\
	 \widetilde{\gamma}_{EB}\neq 0 \Rightarrow f_1(\Delta)&=\infty \leftrightarrow \Delta=4+2k  \ ,
\end{align}
with $k$ a non-negative integer.

\subsubsection*{Results for \texorpdfstring{$\ell \geq 2$}{l>=2}}

This is the most general situation where all nine form factors enter the game. The constraint now takes the form of a linear relation between vectors, i.e.\ a matrix relation,
\begin{align}
	\underbrace{\begin{pmatrix}
		\widetilde{\gamma}_{Bk}^{1} \\
		\widetilde{\gamma}_{Bk}^{2}
	\end{pmatrix}}_{\widetilde{\gamma}_{Bk}}&=\overbrace{\underbrace{\frac{\tan(\frac{\pi(\Delta+\ell)}{2})\left(\frac{\Delta-\ell-3}{2}\right)_{\ell+1}}{(\Delta+\ell)\left(\frac{\Delta-\ell-2}{2}\right)_{\ell+1}}}_{f_{\ell}(\Delta)}\begin{pmatrix}
		\Delta-1 & 2+\ell-\Delta \\
		1-\ell -\Delta & 2(\Delta+\ell-1)
	\end{pmatrix}}^{A_\ell(\Delta)}\underbrace{\begin{pmatrix}
		\gamma_{Ek}^{1} \\
		\gamma_{Ek}^{2}
	\end{pmatrix}}_{\gamma_{Ek}} \ ,
\end{align}
and we will now use matrix notation. The tangent arises through a special ratio of $\Gamma$-functions. This matrix has no null vector, but it may wholly be zero or infinite provided
\begin{align}
	f_\ell(\Delta) &= 0 \; \, \leftrightarrow \Delta=4+\ell+2k \ , \\
	f_{\ell}(\Delta)&=\infty \leftrightarrow \Delta=5+\ell+2k \ .
\end{align}
In these cases, one of the vectors of OPE coefficients must vanish.

To investigate the relation between the different OPE coefficients of $E$ and $B$, we can use a chain of relations previously highlighted in the scalar case, noting that the S-duality argument still holds. There are in principle four different vectors when a $B$ field is present, $\gamma_{BB},\widetilde{\gamma}_{BB}$, $\gamma_{BE},\widetilde{\gamma}_{BE}$. They are related using (in matrix notation)
 \begin{align}
 	\widetilde{\gamma}_{BB}&= (-1)^{\ell+1}\widetilde{\gamma}_{EE} \ , \\
 	\widetilde{\gamma}_{BE}&= A_{\ell}\gamma_{EE} \ , \\
 	\gamma_{BE}&= -A_\ell^{-1}\widetilde{\gamma}_{EE} \ , \\
 	\gamma_{BB}&= \underbrace{(-1)^{\ell}A_\ell^{-1}\sigma_1 A_\ell}_{M}\gamma_{EE} \ .
 \end{align}
 
 It is convenient to differentiate even and odd spin. For odd spin, $\gamma_{EE}=0$, and the discussion is similar to what happened for $\ell=1$.
 %with the first constraint
 %setting $\widetilde{\gamma}_{BB}=0$. 
 %\chris{having trouble here}
 The $\widetilde \gamma_{EB}$ and $\gamma_{BB}$ 
 coefficients are then also sent to zero as well, provided
 $0 < f_\ell(\Delta) < \infty$.  
 %(In the special cases where $f_\ell(\Delta)$ 
 %vanishes or diverges, $\widetilde \gamma_{EB}$ is unconstrained.)
 For even spin, provided $A_\ell$ is invertible, these relations hold, reducing the system to the data of the electric channel. We can note that the matrix $M$ satisfies $M^2=1$, and one can pick a basis of OPE coefficients which diagonalise it.
 
 \vskip 0.1in
 
 In conclusion, we have derived a set of relations between the OPE coefficients of the electric and magnetic fields. We should keep in mind that the $E$--$B$ basis of currents is not orthonormal, giving rise to a non-trivial mixing problem which inputs the values of $(\chi_e,\chi_o)$ into these relations. For each point on the conformal manifold of our theory, these constraints form a non-trivial set of equations relating the OPE coefficients of one current to the other.

 \section{The Bootstrap Problem}
 \label{sec:bootstrap}
 
 While in future work we intend to impose directly the regularity constraints on the three-point functions and bootstrap the full $(E, B)$ system, in this work we will content ourselves by looking at a bootstrap of a single current, which could
 be $E$ or $B$, and the much weaker constraints that come from assuming the presence of a displacement operator and a gap in the spectrum of spin two operators.  We furthermore restrict to theories with a parity symmetry to further constrain the theory.
 
 In general, in the OPE of two currents, we expect there to be a tower of symmetric traceless tensor operators $T_\ell$.
 The four-point function can be decomposed into a sum over the conformal blocks associated with these operators
 and their OPE coefficients.  Schematically, we find crossing equations of the form
 \begin{eqnarray}
  0 &=& \sum_{\Delta} \gamma_{0,\Delta} V_{0, \Delta} \gamma_{0,\Delta} +  \sum_{\Delta }  
 \widetilde \gamma_{0,\Delta} \widetilde V_{0, \Delta} \widetilde \gamma_{0,\Delta} \\
  &&
+   \sum_{\ell = 2}^\infty \sum_{\Delta }  \widetilde \gamma_{\ell, \Delta} \widetilde V_{\ell, \Delta} 
\widetilde \gamma_{\ell,\Delta} +
\sum_{\ell \in 2 {\mathbb Z}^+} \sum_\Delta ( \gamma^{1}_{\ell, \Delta}  \; \gamma^{2}_{\ell, \Delta} ) V_{\ell, \Delta} \left(
 \begin{array}{c}
  \gamma^{1}_{\ell, \Delta}  \\
  \gamma^{2}_{\ell, \Delta}  
 \end{array}
 \right) \ .
  \nonumber
 \end{eqnarray}
 In this notation for the OPE coefficients $\gamma_{\ell, \Delta}$ and conformal blocks $V_{\ell, \Delta}$, the presence of a tilde indicates odd parity and its absence even parity.  
 For identical currents, no operator of spin one is exchanged, but all other non-negative spins are allowed.  
Of course these are tensorial objects, and in fact in appendix  \ref{sec:numerics} we detail how this single schematic crossing equation is actually 5 distinct crossing equations for the different tensor components.

\subsection{Finding Gaps}

As a sanity check, we first reproduce an exclusion plot from 
the original 
four current bootstrap paper \cite{Dymarsky:2017xzb}.   The setup of 
ref.\ \cite{Dymarsky:2017xzb} has a couple of major technical differences.
The first is that the conformal blocks of ref.\ \cite{Dymarsky:2017xzb} were
generated expressly for that work while ours were generated using 
Blocks\_3d \cite{Erramilli:2020rlr}.  A second is that the authors expressed the 
crossing equations in terms of the $(u,v)$ cross ratios while our system
works with $z$ and $\bar z$ where $z \bar z = u$ and $(1-z)(1-\bar z) = v$.

\begin{figure}[h!]
\begin{center}
\includegraphics[width=0.8\linewidth]{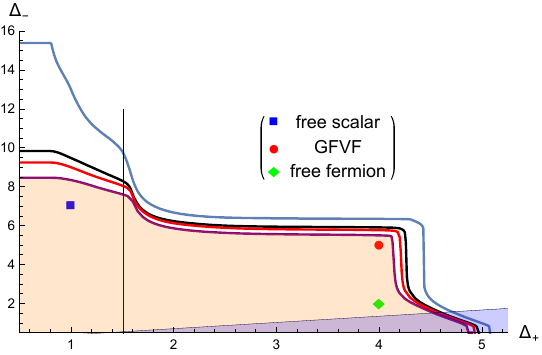}
\end{center}
\caption{
 The exclusion plot  for the four current bootstrap with gaps for the even and odd parity scalars on the axes. 
 The allowed region lies below the curves.
 Our plot is very similar to fig.\ 5 of \cite{Dymarsky:2017xzb} and gives us confidence that the numerics is working.
 The curves correspond to exclusion runs with $\Lambda = 11$, 15, 17, and 25.  The points correspond to known theories: the free scalar, the generalized free vector field, and the free fermion.  The vertical line at $\Delta_+ = 1.5117$ corresponds to the lightest even parity operator in the O(2) model.
 The blue wedge at the bottom is scraped from the standard O(2) CFT bootstrap of four scalars.
 \label{fig:exclusionplot}
}
\end{figure}

 In this exclusion plot, the authors \cite{Dymarsky:2017xzb} ask SDPB  the question, does a CFT exist
with no even parity scalar below $\Delta_+$ and no odd parity scalar below $\Delta_-$.  
Our (very similar) version of their plot is figure \ref{fig:exclusionplot}.  The different
curves correspond to increasing the number of linear functionals that SDPB uses in
examining the crossing constraints.  The number of linear functionals scales
roughly as the square of the parameter $\Lambda$, which in turn is the
maximal derivative order used in constructing these functionals.
As the number of functionals increase, the bounds
become tighter, and the allowed CFTs are below the curves.

As expected, the free scalar, 
free fermion, and GFVF are all allowed.  (Some details about these trivial theories
are reviewed in appendix \ref{sec:free}.)  The one obvious corner
or cusp in the plot seems to be associated with the GFVF.  It is possible that if 
$\Lambda$ were increased further, there may also be corners associated with the 
free scalar and the O(2) model.  For the O(2) model, the smallest odd parity
scalar has not yet been determined; the vertical line corresponds
to the placement of the lowest even parity scalar.

In the context of  boundary CFT, the vertical section of the curves near $\Delta_+ = 4$ is 
noteable.  All boundary theories with a four dimensional bulk
should have a displacement operator of scaling dimension 4.
It seems plausible from this plot that indeed 
quite generally all CFTs with a global U(1) symmetry must have an even parity scalar gap of
at most four.  (The foot extending to larger values of $\Delta_+$ can be ruled out by
bootstrap of four scalars in the O(2) model \cite{Kos:2015mba}.)

\begin{figure}[h!]
\begin{center}
\includegraphics[width=0.7\linewidth]{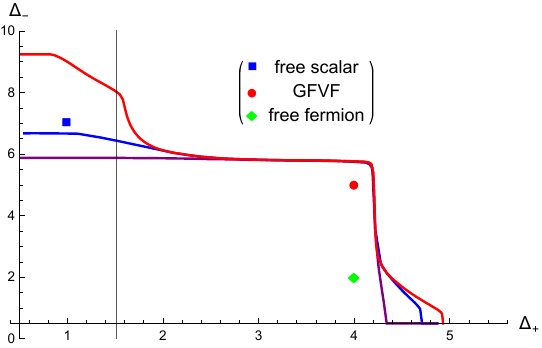}
\end{center}
\caption{
 Three different runs with $\Lambda = 17$ and increasing values of the spin 2 gap.  From top to bottom, the values are 0, 1/2, and 1.  The plot is consistent with the hypothesis that at large enough spin gap, the only theory left is the GFVF and
 theories trivially related.
\label{fig:spin2exclusion}
}
\end{figure}

As the boundary theory should not have its own stress tensor, which is a spin two operator saturating the unitarity bound $\Delta_2 = 3$, we next ask what happens to this exclusion plot fig.\ \ref{fig:exclusionplot} 
if we introduce a spin 2 gap.  The results, which are new, are shown in fig.\ \ref{fig:spin2exclusion}.  
As expected, the free scalar, which has a stress tensor, is excluded once the gap becomes large enough.
(Computing the exclusion plot with a smaller value of $\Lambda$, the free scalar is  excluded at a
larger value of the gap.)  The GFVF on the other hand does not have a stress tensor but does have a spin two operator of dimension 4, namely $V_{\mu}V_{\nu} - \frac{1}{3} \delta_{\mu\nu}V^2$.  The GFVF
survives even when the spin 2 gap is increased to one.  The free fermion appears to be allowed according to this plot
but presumably only because it is protected inside the convex hull of the region fixed by the presence of the GFVF.

\begin{figure}[h!]
\begin{center}
\includegraphics[width=0.7\linewidth]{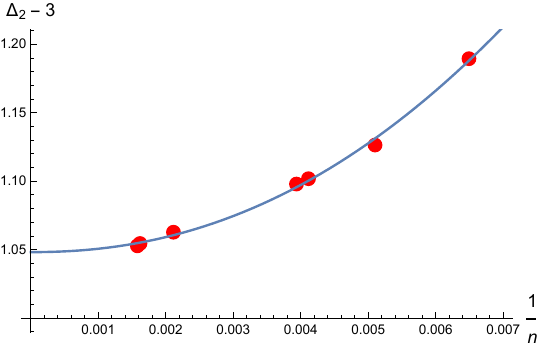}
\end{center}
\caption{
The maximal allowed spin 2 gap vs.\ the inverse number of linear functionals used in SDPB.  Note that
the number of linear functionals scales roughly as the derivative order squared, $\Lambda^2$.  
We also have provided a fit to the data of the form $\Delta_2 - 3 = a n^\alpha + b$ where the best fit parameters for
these seven data points were $a=7530$, $b=1.05$ and $\alpha = -2.16$, suggesting the convergence is quadratic
in the number of functionals and that there are no CFTs for a spin 2 gap a little bit bigger than one.  
\label{fig:spin2gap}
}
\end{figure}

If we continue to increase the spin 2 gap, eventually all CFTs are ruled out, even if no gap assumptions are made for the scalar sector, by which we mean the scalar dimensions can saturate the unitarity bound of one half.  What this critical value of the spin 2 gap is, depends on how we set the number of linear functionals or equivalently $\Lambda$. 
In fig.\ \ref{fig:spin2gap}, we show the maximal allowed value for the spin 2 gap as a function of the inverse number of linear functionals.  In this way, we can try to fit the points to a curve and extrapolate the result to the infinite
$\Lambda$ limit.  The result is that there are no CFTs once the spin 2 gap exceeds 1.05, which is suspiciously close to one.  We are tempted to speculate that in fact the true upper bound is one, that the unique CFT with a spin 2 gap of one is the GFVF or something constructed from it in a trivial way, for example by taking products of the GFVF and orbifolding, and that there is no CFT with a larger spin 2 gap.

\subsection{OPE Maximisation and the Displacement}

\begin{figure}[h!]
\begin{center}
\includegraphics[width=0.75\linewidth]{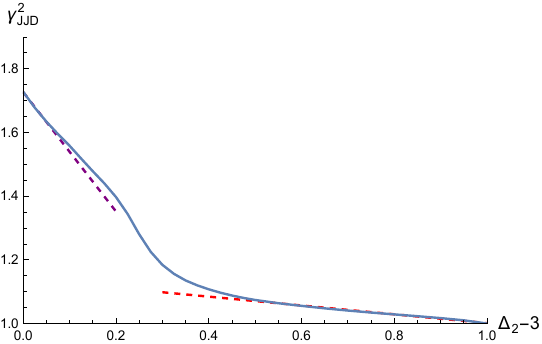}
\end{center}
\caption{
A bound on $\gamma_{JJD}^2$ as a function of the spin 2 gap.  Allowed theories must lie below the curve.
The OPE coefficient has been normalized such that
for a gap of one, $\gamma_{JJD}^2 = 1$.  The value of $\gamma^2_{JJD}$ when the gap is zero is approximately 1.728.  
The slope of the line at that point is $-1.873$.  The slope of the line when the gap is one is $-0.139$.  Dashed lines with
these slopes have been plotted as a guide to the eye.
The curve was computed by using linear extrapolation for runs with $\Lambda = 11$, 17 and 25.  The bounds from these three runs sit very nearly on a line when plotted against the inverse number of linear functionals input to SDPB (117, 243, and 473 respectively).
\label{fig:OPEplot}
}
\end{figure}

\begin{figure}[h!]
\begin{center}
\includegraphics[width=0.55\linewidth]{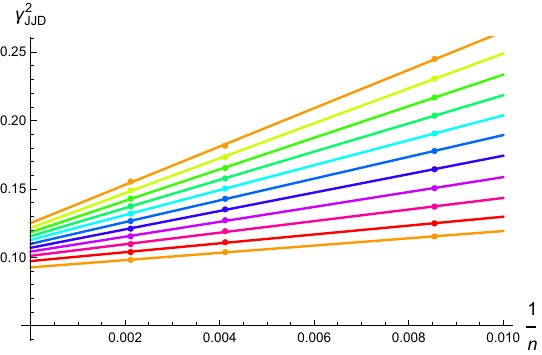}
\end{center}
\caption{
The linear extrapolation used to compute the OPE bound in fig.\ \ref{fig:OPEplot}.  
For each final point in fig.\ \ref{fig:OPEplot}, SDPB was run with $\Lambda=11$, 17 and 25, corresponding to 117, 243, and 473 independent linear functionals.  The $x$-axis of the plot is the inverse number $1/n$ of linear functionals.  From top to bottom, the lines
correspond to spin 2 gaps from zero to one.
\label{fig:extrapolation}
}
\end{figure}

In addition to checking whether CFTs with certain gaps in the operator spectrum are allowed, SDPB is also useful for computing bounds on OPE coefficients.  In this next plot, we examine the OPE coefficient of two currents with 
an even parity, dimension four operator, which could be the displacement operator in our boundary CFT.
An upper bound for this coefficient is plotted in fig.\ \ref{fig:OPEplot}. We normalised with respect to the result obtained when the spin two 
gap becomes equal to one, which should correspond to the value in the GFVF. In this plot, we have assumed that the scalar operators satisfy the unitarity bound $\Delta_\pm \geq \frac{1}{2}$.  
Once the spin 2 gap reaches one, our hypothesis is that the only theory left is the GFVF.  
Note that fig.\ \ref{fig:OPEplot} was computed by a linear extrapolation of the results for runs at three lower values of 
$\Lambda$.  The extrapolation is pictured in fig.\ \ref{fig:extrapolation}.

\begin{figure}[h!]
\begin{center}
\includegraphics[width=0.7\linewidth]{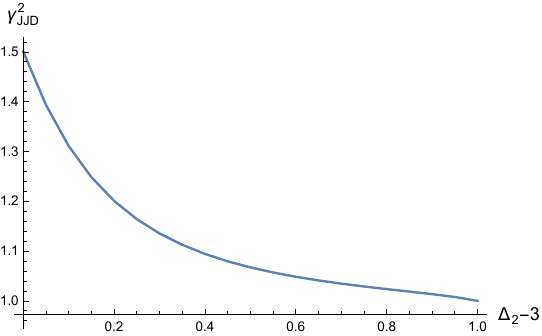}
\end{center}
\caption{A bound on $\gamma^2_{JJD}$ as a function of the spin 2 gap but now with gap assumptions $\Delta_+ \geq 4$ and $\Delta_- \geq 2$ designed to pick out the free fermion theory when $\Delta_2 = 3$.  No extrapolation was used to make this plot, which corresponds to an SDPB run with $\Lambda = 25$.  The OPE coefficient has been normalized such that $\gamma_{JJD}^2 =1$ when $\Delta_2 = 4$.  
The value at $\Delta_2 = 3$ (and in fact the whole plot) changes little as a function of $\Lambda$, increasing from 1.49957 at $\Lambda =11$ to 1.50067 at $\Lambda = 25$.}
\label{fig:ratioOPEplot}
\end{figure}

To provide some evidence that the theory with $\Delta_2 - 3 \geq 1$ that is saturating the OPE bounds is the GFVF, we take a another section through the OPE coefficient space.  This time, we gradually increase the spin 2 gap from zero but assuming that there are no even parity scalars with $\Delta_+ < 4$ and no odd parity scalars with dimension $\Delta_- < 2$.  These gap assumptions are designed to allow the free fermion theory when the spin 2 gap is vanishing.  The corresponding plot is shown as fig.\ \ref{fig:ratioOPEplot}.
The plot is again normalized such that the OPE coefficient bound at $\Delta_2 = 4$ is equal to one.
The ratio between the value at $\Delta_2 = 3$ and the value at $\Delta_2 = 4$ is very close to $\frac{3}{2}$, which is the ratio between the free fermion and the GFVF calculated in appendix \ref{sec:free}.  (For an SDPB run
with $\Lambda = 25$, the precise value obtained was 1.50067.)
It should also be emphasized that although the bound computed by SDPB at $\Delta_2 = 4$ changes slightly as a function of $\Lambda$, it is insensitive to the choice of gap assumption, $\Delta_\pm \geq \frac{1}{2}$ or
$(\Delta_+ \geq 4, \Delta_- \geq 2)$.

Our last plot is the one most constrained by the notion that there is a bulk Maxwell field coupled to a CFT with a global U(1) symmetry.  Using the Ward identity of \cite{DiPietro:2019hqe}, we relate the OPE coefficient $\gamma_{JJD}$
 to the two-point functions of the currents and the displacement operator:
 \begin{equation}
 \gamma_{JJD}= \frac{2}{\pi^2}\tau_{JJ} +\frac{\kappa}{\pi^2}\left(\frac{C_{D}}{C_{D}^{free}}-1\right) \ , 
 \end{equation}
 where $J$ can be either $E$ or $B$.  This result follows directly from (\ref{gammaJJD}), and the derivation is reviewed in appendix \ref{sec:Ward}.
 
\begin{figure}
\begin{center}
\includegraphics[width=3.5in]{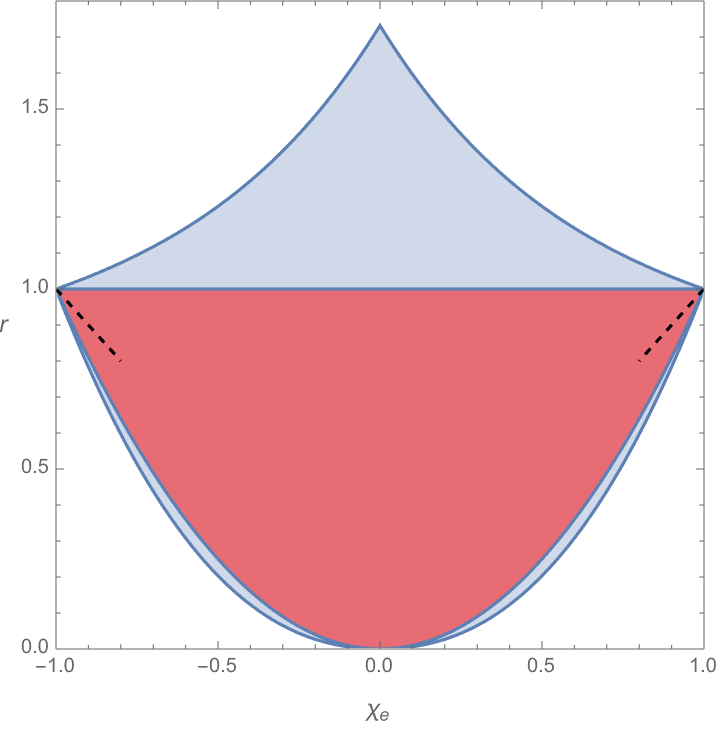}
\end{center}
\caption{
The allowed region in the $r = C_D / C_D^{free}$ vs.\ $\chi_e$ plane.  The larger region assumes 
$\gamma_{BB}^2 / \tau_{BB}^2 C_D$ and $\gamma_{EE}^2 / \tau_{EE}^2 C_D$ are bounded above by
$1.728 \times \frac{1}{6}$.  For the red region, the upper bound is instead the GFVF result $\frac{1}{6}$.  The black lines in the corners correspond to 
perturbation theory.  Note the plot gives a rigorous upper bound on the $b_2$ boundary anomaly coefficient.
\label{fig:DDplot}
}
\end{figure}

The upper bound on $\gamma_{JJD}$ then translates into bounds on $\tau_{EE}$, $\tau_{BB}$, and $C_D$, pictured in fig.\ \ref{fig:DDplot}.    As the spin 2 gap is increased, the blue region shrinks down to the red region.  
(The results of perturbation theory reviewed in section \ref{sec:perturbativeperspective} are shown as the dashed lines, which lie comfortably inside the allowed region.)  
The upper bound of 1.728 in fig. \ref{fig:OPEplot} translates into an upper bound on the displacement two-point  function, namely that $C_D \leq 1.728 \, C_D^{free}$.

This result is doubly interesting as $C_D$ is well known to control 
boundary contributions to the anomalous trace of the stress tensor \cite{Herzog:2017vj,Herzog:2017kkj}, as we 
review in the discussion.

\section{Discussion}
\label{sec:discussion}

To conclude, we discuss three facets of this work.
First, we discuss the two-point function sum rule and a possible interpretation that may shed some light on the puzzle surrounding the optical conductivity of graphene.
Second, we further develop the bound on the displacement two-point function and its relation to conformal anomalies.  Finally, we describe the next steps.

There is a curious relation between boundary QED and graphene.  Effective field theory models of graphene typically involve massless Dirac fermions with a linear dispersion relation $E = v p$, where the velocity $v \approx c / 300$ is small compared to that of light, small enough that magnetic interactions can be effectively ignored. 
The speed $v$ however develops a negative beta function, and thus becomes larger at low energies.  Researchers
have argued \cite{Gonzalez:1993uz} that boundary QED can be thought of as the ultimate IR fixed point of graphene, where $v \to  c$ and magnetic interactions are restored.  

The optical conductivity puzzle starts with a calculation in the free field limit of graphene.  Using the current $J_\mu = e \bar \psi \gamma_\mu \psi$ with two-component fermions, 
the conductivity can be calculated from Wick's Theorem to be
\begin{equation}
\sigma = \frac{N_f e^2}{16} \ .
\end{equation}
For graphene, the expectation is that $N_f = 4$, and remarkably this result matches experiment at the percent level for electric field frequencies in the optical range \cite{Mak_2008,NairBlakeGrigorenkoetal} .
The result follows from a Kubo formula
\begin{equation}
\sigma = \lim_{{\bf k} \to 0} \frac{1}{i \omega} G_{xx} (\omega, {\bf k})  \ ,
\end{equation}
where $k = (\omega, {\bf k})$ and $G_{\mu\nu}(k)$ is the Fourier transform of the current two-point function
\begin{equation}
G_{\mu\nu}(k) = \int d^3 x \, e^{-i k \cdot x} \langle J_{\mu}(x) J_\nu(0) \rangle \ .
\end{equation}
This conductivity is that of the zero temperature limit, which in practice means that the ratio $\omega/ T$ should be taken large.  In other words, one is measuring the current response of graphene to an oscillating electric field, for example that in a beam of light.  

The puzzle is that this free field calculation is suspect; the effective fine structure constant in graphene, because of the relatively low value of $v$, is order one,
\begin{equation}
\alpha_g = \frac{4 \pi}{e^2 v}  \approx 2.2 \ ,
\end{equation}
suggesting Coulomb interactions should be important.  Researchers have tried to address these concerns perturbatively by calculating the first order correction to the conductivity.  In an expansion in $\alpha_g$, the first order correction turns out to be nummerically small and positive \cite{Mishchenko_2008},
\begin{equation}
\sigma = \frac{e^2}{4} \left(1 + \frac{19-6\pi}{12} \alpha_g + O(\alpha_g^2) \right) \ .
\end{equation}
A similar calculation has also been performed in boundary QED, where it is also small and positive \cite{Teber_2012,Kotikov_2013}
\begin{equation}
\sigma = \frac{e^2}{4} \left( 1 + \frac{92-9 \pi^2}{18 \pi} \alpha + O(\alpha^2) \right) \ .
\end{equation}

Of course these are just perturbative results, and one would like some additional non-perturbative input.  
A result of \cite{DiPietro:2019hqe} that we rederived in section \ref{sec:FF} 
can be rephrased as a sum rule for the conductivities
of the two currents, one of which is the charge current just discussed,\footnote{%
 Earlier, we used a different normalization for the current two-point functions.  
 To make the translation, there is a factor of $\pi^2/2$ that comes from the Fourier transform
 (see footnote \ref{footnote:normalization})
 and an additional factor of $g^2$ that is removed in translating the electric current $E$ to 
 a charge current $J$ which, unlike ${\mathcal J}$ used previously, includes a factor of the gauge
 coupling in its definition.
 Note this sum rule was also found in a supersymmetric context from localization \cite{KumarGupta:2019nay}.  
}
\begin{equation}
\sigma + \sigma' = 1 \ .
\end{equation}
As both conductivities need to be positive by reflection positivity, we learn that
$\sigma \leq 1$.  A further consequence is that some of the higher order perturbative corrections to $\sigma$ 
must be negative.  
If the theory is self-dual under swapping the two currents, it follows also that $\sigma = \sigma' = \frac{1}{2}$
at the self-dual point, generalizing a result of \cite{Hsiao:2017lch,Hsiao:2018fsc}.  

Regarding the  trace anomaly, famously four dimensional CFTs have two scheme independent
contributions to the trace of the stress tensor proportional to the Euler density and the Weyl curvature squared, with coefficients  called respectively $a$ and $c$.  Less well known is that in the 
presence of a boundary, there are two additional terms.   (That these are the only Wess-Zumino consistent terms was shown in \cite{Herzog:2015ioa}.)  Schematically, the full trace anomaly takes the form
\begin{equation}
T^\mu_\mu = \frac{1}{16\pi^2} \left( -a \, {\mathcal E} + c\,  W^2 \right) + \frac{\delta(x_\perp)}{16\pi^2} \left( a \, {\mathcal E}_{\rm bry} -b_1 \tr \hat K^3  - b_2 \, h^{\alpha \gamma} \hat K^{\beta \delta} W_{\alpha \beta \gamma \delta} \right) \ ,
\end{equation}
where ${\mathcal E}$ is the Euler density and ${\mathcal E}_{\rm bry}$ its boundary contribution, $W_{\mu\nu\lambda\rho}$ the Weyl curvature, $K_{\mu\nu}$ the extrinsic curvature of the boundary and $\hat K_{\mu\nu}$ its traceless part.
It is known from \cite{Herzog:2017vj,Herzog:2017kkj} that $b_2$ is proportional to the displacement operator two-point function coefficient $c_D$ and
$b_1$ to the displacement operator three-point function coefficient, essentially because for a slightly deformed planar boundary, the extrinsic curvature is proportional to the Hessian of the boundary's position.  
For many years, people have speculated on possible constraints on these coefficients.  
It is immediately clear from reflection positivity that $b_2 \geq 0$ \cite{Herzog:2017kkj}.  Fursaev \cite{Fursaev:2015wpa}
noted a relationship between $b_2$ and the bulk coefficient $c$ for free theories, $b_2 = 8 c$, while one of us \cite{Herzog:2017kkj}  demonstrated that $b_2$ changes as a function of the marginal coupling $g$ in boundary QED while $c$ (which is determined by its value for free Maxwell theory) does not.
Here at last we have
a strict upper bound on $b_2$ through our result that $C_{D} \leq 1.728\,  C_D^{free}$, at least for boundary CFTs with a global U(1) that are coupled to free Maxwell theory.   
 We find that
\[
b_2 \leq 13.8\,  c \ .
\]
It will be interesting to see if this bound applies more generally and also to see if it can be tightened further.

The coefficient $b_1$ depends on a displacement three-point function and is not direclty accessible through this four current bootstrap.  One analytic result that is known is a sum rule that relates $b_1$ to $b_2$ and also all the spin two operators that appear in the OPE of two displacement operators \cite{Herzog:2021spv}.  It would be interesting to try to constrain $\langle DDD \rangle$ by bootstrapping a system of four scalar operators of dimension four, assuming a spin 2 gap above the unitarity bound in order to restrict to theories without a boundary stress tensor.  Such a project should be interesting and probably much simpler than the higher spin bootstrap performed here and the more complicated bootstrap project we have in mind next.

As part two of this project, we intend to look at the full system of four current correlation functions, involving both the
electric $E$ and magnetic $B$ currents, imposing the constraints on the OPE coefficients we derived in section \ref{sec:threepoint}.  We hope to be able to follow the changes to the free fermion and free scalar once they
are coupled to the bulk Maxwell field, using the spin 2 gap as a proxy for the coupling strength.  For small
values of the coupling, the results are known in perturbation theory, which we can then hopefully match to our bootstrap results.  With confidence in the bootstrap established, we can then use the bootstrap in a nonperturbative regime to compute anomalous dimensions and OPE coefficients.  That, anyway, is the plan.

\section*{Acknowledgements}
We would like to thank Petr Kravchuk, Marten Reehorst, and Andy Stergiou for patiently answering endless questions about the numerical conformal bootstrap.  Without their help, we would still be wondering how to format the input files for SDPB.  We would also like to thank Conor Behan, Dalimil Maz\'a\v{c}, Silviu Pufu, and Slava Rychkov for discussion.  VS is grateful to Kavli IPMU for hospitality during time when part of this work was carried out.
This work was funded in part by  a Wolfson Fellowship from the Royal Society
  and by the U.K.\ Science \& Technology Facilities Council Grant ST/P000258/1.
  Computations were performed on the HPC CREATE KCL cluster, COSMA8 at the University of Durham, and ARCHER2 at Edinburgh.

\appendix

 \section{Perimeter Implies Free}
 \label{sec:perimeterimpliesfree}
 
 Consider a theory for which the $\langle F_{\mu\nu} F_{\lambda \rho} \rangle$ two-point
 function lies on the perimeter of the disk $\chi_e^2 + \chi_o^2 \leq 1$.
 We present an argument that it must correspond to a free theory.
 The argument takes advantage of the OPE constraints derived in section \ref{sec:OPEconstraints}.  
 
 As we saw in section \ref{sec:FF}, on the perimeter of the disk,  $E\parallel B$ are the same operator. 
 Let us write then
 \begin{align}
 B = \alpha E \ ,
 \end{align}
 for $\alpha$ some real number.  
 As a result, the OPE coefficients must be related via $\gamma_{BB} = \alpha \gamma_{EB} = \alpha^2 \gamma_{EE}$
 and similarly for the odd parity coefficients $\widetilde \gamma$.    These relations are compatible with the constraints worked out in section \ref{sec:OPEconstraints}  only in 
 special circumstances.
 
% We are only able to probe the part of the boundary spectrum which is measured by $E$ and $B$, so let us take 
%%
%\begin{align}
%	E &= \abs{\alpha} J &  B &= \abs{\beta}J & \lambda &=\frac{\abs{\beta}}{\abs{\alpha}}\geq 0
%\end{align}
%%
%We can now investigate the OPE constraint for each of the fields contained in $J(x)J(y)$. The goal is to find the scaling dimensions of the allowed operators 

\subsubsection*{Scalar spectrum} Let us first consider $f(\Delta)\neq 0,\infty$. 
The relation (\ref{eq:gammascalar}) implies $\gamma_{EE} = -\gamma_{BB}$ which in general contradicts with $\gamma_{BB} = \alpha^2 \gamma_{EE}$. 
To get non-trivial OPE coefficients, 
we must set $f(\Delta)=0,\infty$, which in turn severely restricts the spectrum
and sets half of the OPE coefficients to zero.
%The same argument applies to $\widetilde{\gamma}$. 
In summary, we have the two cases
%
%
%Using $f(\Delta)\gamma_{BE}=i\widetilde{\gamma}_{EE}$, we have $f(\Delta)\gamma =\lambda i\widetilde{\gamma}$. This gives 
%
\begin{align}
	f(\Delta)=0 \leftrightarrow \Delta = 4+2k &\rightarrow \gamma \neq 0, \widetilde{\gamma}=0 \ , \\
	f(\Delta)=\infty \leftrightarrow \Delta = 5+2k &\rightarrow \gamma = 0, \widetilde{\gamma}\neq 0  \ ,
\end{align}
for $k$ a non-negative integer.
These two towers of operators are precisely the ones that appear in the scalar sector of a GFVF. 
In the OPE of a current $V$ with itself one finds
$[V\cdot V]_{k,0}$ with dimension $4+2k$ and $[\partial\cdot V \wedge V]_{k,0}$ with dimension $5+2k$
as mentioned in section \ref{sec:perturbativeperspective}.  

\subsubsection*{Vector Spectrum}
Bose symmetry allows only for $\gamma_{EB}$ and $\widetilde \gamma_{EB}$ to be nonzero.  Moreover, 
$\gamma_{EB} \neq 0$ only for the tower of operators $\Delta = 5 + 2k$ while 
$\widetilde \gamma_{EB} \neq 0$ only for the tower of operators $\Delta = 4 + 2k$.  The only operators which could potentially enter have the dimension and parity of the operators 
$[V\cdot V]_{k,1}$ and $[V \wedge V]_{k,0}$ in the free theory. 

In fact things are even more constraining.  Since $\gamma_{EB}$ and $\widetilde \gamma_{EB}$ are related to the other
coefficients $\gamma_{EE}$, $\gamma_{BB}$, $\widetilde \gamma_{EE}$ and $\widetilde \gamma_{BB}$ through factors of
$\alpha$, we conclude in general that all of these OPE coefficients must vanish for consistency.  

\subsubsection*{Even Spin Tensor Spectrum} 
We look at $\ell=2n\geq 2$ and begin by assuming $f_{\ell}(\Delta)\neq 0,\infty$. 
Consider the relation $\widetilde{\gamma}_{BB}=(-1)^{\ell+1}\widetilde{\gamma}_{EE}$ 
(which assumes $f_{\ell}(\Delta)\neq 0,\infty$).  This relation is in contradiction with
$\widetilde \gamma_{BB} = \alpha^2 \widetilde \gamma_{EE}$ except if both coefficients vanish.

We also have $\gamma_{BE}=-A_\ell^{-1}\widetilde{\gamma}_{EE}$ and $\widetilde \gamma_{BE} = A_\ell \gamma_{EE}$, which means $\alpha \gamma_{EE} = - A_\ell^{-1} \widetilde{\gamma}_{EE}$ and $\alpha \widetilde \gamma_{EE} = A_\ell \gamma_{EE}$. If $f_\ell(\Delta)\neq 0,\infty$, vanishing of $\widetilde \gamma_{EE}$ sets $\gamma_{EE} =0$. It then follows all the OPE coefficients vanish.

The exceptions are the cases where $f_\ell(\Delta)$ vanishes or diverges. If $f_\ell(\Delta)=\infty$, the even parity coefficients vanish $\gamma =0$. If $f_{\ell}(\Delta)=0$, the odd parity ones disappear $\widetilde{\gamma}=0$. But in each case the opposite parity OPE coefficients survive. 
The end result  for even spins is that 
\begin{align}
	f_{2n}(\Delta) &= 0 \; \, \leftrightarrow \Delta=4+2n+2k \Rightarrow \gamma \neq 0  \ , \\
	f_{2n}(\Delta)&=\infty \leftrightarrow \Delta=5+2n+2k \Rightarrow \widetilde{\gamma} \neq 0 \ , 
\end{align}
which as in the previous cases, reduces to the dimensions and parity selection rules of the free field normal ordered operators $[V\cdot V]_{k,2n}$ and $[\partial\cdot V \wedge V]_{k,2n}$.

\subsubsection*{Odd Spin Tensor Spectrum}

We now pick $\ell = 2n+1$. For these operators we know from the crossing selection rules that $\gamma_{EE} = 0 = \gamma_{BB}$. The remaining rules 
$\gamma_{BE}=-A_\ell^{-1}\widetilde{\gamma}_{EE}$ and $\widetilde \gamma_{BE} = A_\ell \gamma_{EE}$
 imply $\alpha \gamma_{EE}=-A_\ell^{-1}\widetilde{\gamma}_{EE}$ and $\alpha \widetilde \gamma_{EE} = A_\ell \gamma_{EE}$. The conditions allow for a non-trivial $\widetilde{\gamma}_{EE}$ only if $A_\ell^{-1}=0$, and so if $f_{\ell}(\Delta)=\infty$. The allowed set of operators is then 
\begin{align}
	f_{2n+1}(\Delta)=0 \leftrightarrow \Delta = 4+2k+(2n+1) &\rightarrow  \widetilde{\gamma}\neq 0 \ ,  
\end{align}
which correspond to the dimension and parity of the tower $[\partial\cdot V \wedge V]_{k,2n+1}$.

\subsubsection*{Summary and Implications}

We have seen that our OPE relation implies that if $E\parallel B$, the OPE of the boundary currents can be summarised using the dimension and parity corresponding to the primaries 
\begin{align}
	[V\cdot V]_{n,2k}, \, [\partial\cdot V \wedge V]_{n,2+k}, \, k \in \mathbb{N} \ .
\end{align}
It is a standard result, proved \cite{Lauria:2021ut} and used extensively in \cite{Herzog:2022jqv}, that if two fields possess the OPE content of a generalised free field theory, their correlation functions are gaussian, and so they are genuine free fields. The proof follows by noting that all the discontinuities of the correlation functions are encoded in the two-point function.  Inductively, a generic $(n+2)$-point function can be reduced to its discontinuity on branch-cuts, reducing it to a product of a two-point function and an $n$-point correlator, thus ultimately reducing it to a sum over products of two-point functions.

We have established that the ring is solely made up of free theories, which are precisely those we found from the Lagrangian picture.

\section{Details About the Numerics}
\label{sec:numerics}

Our numerical bootstrap pipeline has three major pieces.  The first and last piece consist of pre-existing software: Blocks\_3d \cite{Erramilli:2020rlr} and SDPB \cite{sdpb,Landry:2019qug}, while the third was purpose built for this project, adapting the output of Blocks\_3d for input into SDPB.

Blocks\_3d \cite{Erramilli:2020rlr} approximates conformal blocks for three dimensional conformal field theories to arbitrary precision.  The external and exchanged 
operators can be in arbitrary representations of $SO(3)$ and carry arbitrary conformal dimension.  
Often in numerical bootstrap applications, the conformal dimension of the external operators needs to be adjusted for each run of SDPB.  Our case was simpler in this respect as the external currents are protected operators with conformal dimension equal to two.  More than making up for this simplification however were two additional considerations: the tensor structures that accompany spinning operators and the conservation condition that the currents enjoy.

A limitation of Blocks\_3d is that the conformal blocks do not satisfy conservation conditions for the external operators.  
A major part of the second piece of pipeline that we built was the construction of the conserved blocks from linear combinations of the output of Blocks\_3d.  As the algebra required was nontrivial, we implemented this part of the pipeline in Mathematica, following procedures detailed in \cite{He:2023ewx}. 

The other parts of the second piece of pipeline that we built were a collection of C programs and bash scripts to feed the conserved conformal blocks into SDPB.  While we did not need to adjust the dimension of the external operators for 
each run of SDPB, it was useful to be able adjust the gap assumptions quickly and frequently and also to alter the norm and objective.  
Thus we have some C routines that introduce these gaps into the conserved conformal blocks and compute the norm and objective in specially tailored ways.  

The blocks have to be assembled into crossing equations that are the input for SDPB.  
We have C routines that pack different parts of the SDPB input file: some that are run once for input that does not change from run to run, others that pack the scalar or spin two sector whose gap assumptions change frequently, that need to be run prior to each instance of SDPB.  

We have a C routine that removes
$x/x$ pieces from the conserved conformal blocks.  It turns out that our conserved blocks, which are approximated as ratios of polynomials in $x = \Delta-\Delta_0$, where $\Delta_0$ is the unitarity bound for the exchanged operator,
often have factors of the form $x/x$.  In particular our even parity, even spin exchanged blocks, $J \geq 2$ all have this $x/x$ factor, as does the odd parity $J=2$ block.  It was suggested that for improved behavior of SDPB, these factors should be removed.\footnote{%
Personal communication P.~Kravchuk.
}

Finally, we have many bash scripts, some for setting up the static portion of the input file for SDPB, others for running a single instance of SDPB, others to implement binary search routines to search for where a particular behavior
of SDPB changes.

In what follows we sketch the technical details involved in constructing the conserved conformal blocks and constructing the crossing equations.  Our procedure for constructing the conserved blocks follows closely ref.\ \cite{He:2023ewx} as does our construction of the crossing equations.  (For a more detailed account of how to set up the crossing equations in a closely related case, the stress tensor bootstrap, the reader may wish to consult \cite{Dymarsky:2017yzx} from which
both ref.\ \cite{He:2023ewx} and we draw heavily.)

\subsection{Three-Point Functions: Vector to \texorpdfstring{$q$}{q}-Basis}

To construct the conserved blocks, we begin with the three-point functions.  Our initial ``vector'' 
basis of three-point function tensor
structures was described already in section \ref{sec:JJT}. However, to interface with Blocks\_3d, it is necessary to change to an SO(3) basis.  

Consider a conformal block where external operators with $SO(3)$ spins $j_1$ and $j_2$ merge to form
an exchanged operator with spin $j_3$.  Blocks\_3d uses an SO(3) basis that involves 
states with quantum numbers $|j_1, j_2, j_3, j_{12}, j_{123}, m \rangle$.  
The numbers $j_{12}$ and $j_{123}$ are the principle quantum numbers corresponding to the angular momenta
${\bf J}_1 + {\bf J}_2$ and ${\bf J}_1 + {\bf J}_2 + {\bf J_3}$.
Constructing the conserved conformal
blocks from the output of Blocks\_3d involves several steps.  The first we have already accomplished in section \ref{sec:JJT}, figuring out the restrictions on the allowed tensor structures from conservation.  The next step is
to convert this ``vector''  basis to the SO(3) basis $|j_1, m_1; j_2, m_2; j_3, m_3 \rangle$.  Then we use 
Clebsch-Gordan coefficients twice to go to the final $|j_1, j_2, j_3, j_{12}, j_{123}, m \rangle$ basis.
This procedure tells us how to change the basis of the three-point functions.  Of course, we really want to know
how to reassemble the conformal blocks, but knowing the rules for the three-point function is enough to deduce
what happens to the full conformal block; we perform the change of basis twice, one for each pair of external operators in the conformal block.  

In more detail, the first step is to relate our ``vector'' basis of embedding space structures
to the $|j_1, m_1; j_2, m_2 ; j_3, m_3 \rangle$ states.  Note that by angular momentum conservation, we assume
$\sum_i m_i = 0$.  
We fix the insertions in the three-point function to
$x_1 = (0,0,0)$, $x_2 = (0,0,1)$, and $x_3 = (0, 0, \infty)$. 
A prescription was outlined in appendix F of \cite{Erramilli:2020rlr} (available in the published version). One makes the identifications 
\bal
P_1 &= \left(0 , 0 , 0 , \frac{1}{2} , \frac{1}{2} \right) \ , \; \; \;
P_2 = \left( 0 , 0 , 1, 0 , 1 \right) \ , \; \; \;
P_3 = \left( 0 , 0 , 0 , -\frac{1}{2} , \frac{1}{2} \right) \ , \\
Z_1 &= \left( i\frac{\xi_1^2 + \eta_1^2}{\sqrt{2}} , \frac{\xi_1^2 - \eta_1^2}{\sqrt{2}}, - \sqrt{2} \xi_1 \eta_1 , 0 ,0 \right) \ , \\
Z_2 &=  \left( i\frac{\xi_2^2 + \eta_2^2}{\sqrt{2}} , \frac{\xi_2^2 - \eta_2^2}{\sqrt{2}}, - \sqrt{2} \xi_2 \eta_2 , \sqrt{2} \xi_2 \eta_2 , -\sqrt{2} \xi_2 \eta_2 \right) \ , \\
Z_3 &=\left( i\frac{\xi_3^2 + \eta_3^2}{\sqrt{2}} , \frac{\xi_3^2 - \eta_3^2}{\sqrt{2}}, \sqrt{2} \xi_3 \eta_3, 0 , 0 \right) \ ,
\eal
and  then reads off the SO(3) states by expanding the embedding space structure in terms of $\xi_a$ and $\eta_a$ and using the definition (2.28) of \cite{Erramilli:2020rlr}
\be
|j_1, m_1; j_2, m_2 ; j_3, m_3 \rangle = (-1)^{j_1- j_3 + m_2} 
\prod_{a=1}^3 \sqrt{{2j_a \choose j_a + m_a} } \xi_a^{j_a + m_a} \eta_a^{j_a - m_a} \ .
\ee
Using the notation $|1, m_1; 1, m_2 ; j, m_3 \rangle = [m_1 m_2 m_3]$, we can write down the nine $q$-states that contribute to the three-point functions
\[
[000] , \  [1-10] \, [-110] , \ [10-1]  , \ [-101]  , \ [01-1]  , \ [0-11]  , \ [11-2]  ,  \ [-1-12] \ .
\]
Parity in this basis corresponds to the action of the $\sigma_1$ Pauli matrix which swaps $\xi_a$ and $\eta_a$, and hence
sends $m_a \to -m_a$.  Thus the even and odd parity combinations are symmetric and antisymmetric combinations of pairs of states related by this swap.  Note $[000]$ is its own conjugate, and so we get 5 parity even states and 4 parity odd states in general.  For $l=1$, we keep only the first 7 of these states, yielding 4 parity even and 3 parity odd.  For $l=0$, we keep only the first 3, giving 2 parity even and 1 parity odd state.  
These results match the counting from our ``vector'' basis construction.

Next one applies the Clebsch-Gordan relations twice to move to the basis 
\begin{equation}
|j_1, j_2, j_3, j_{12}, j_{123}, m \rangle = |j_{12},  j_{123} \rangle \ .
\end{equation}
One finds the following matrices which implement a change of basis from 
%$|j_1, j_2, j_3, j_{12},
%
$|j_{12}, j_{123} \rangle$ 
to the ``vector'' basis, ordered as in section \ref{sec:JJT}.
For even parity three-point functions with $j\geq 2$:
\bal
\frac{(-1)^j 2^{\frac{j}{2}}}{\sqrt{3 {{2j \choose j}}}}
\left(
\begin{array}{ccccc}
1 & 0 & - \sqrt{\frac{3 j (j-1)}{4j^2-1}} & \sqrt{\frac{2 j (j+1)}{(2j-1)(2j+3)}} & -\sqrt{\frac{3 (j+1)(j+2)}{(2j+1)(2j+3)}} \\
-1 & \sqrt{\frac{3(j+1)}{2j}} & \sqrt{\frac{3(j-1)(2j+1)}{j(2 j-1)}} & -\sqrt{\frac{(j+1)(2j+3)}{2 j (2j-1)}} & 0 \\
-1 & \sqrt{\frac{3(j+1)}{2j}} & -\sqrt{\frac{3(j-1)(2j+1)}{j(2 j-1)}} & -\sqrt{\frac{(j+1)(2j-3)^2}{2 j (2j-1)(2j+3)}} & \sqrt{\frac{12 (j+1)(j+2)}{(2j+1)(2j+3)}} \\
2 & -\sqrt{\frac{6(j+1)}{j}} & \sqrt{\frac{3 (2j+1)}{j (j-1)(2j-1)}} & \sqrt{\frac{2(j+1)(2j+3)}{j(2j-1)} }& 0 \\
-3 & 0 & 0 & 0 & 0
\end{array}
\right)
\eal
The above acts on $|0, j\rangle$, $|1, j \rangle$, $|2, j-2 \rangle$, $|2, j\rangle$, and $|2, j+2 \rangle$.

\noindent
We have also: Even parity sector with $j=1$,
\bal
\left(\begin{array}{cccc}
-\frac{1}{\sqrt{3}} & 0 & -\frac{2}{\sqrt{15}} & \sqrt{\frac{2}{5}} \\
\frac{1}{\sqrt{3}} & -1 & \sqrt{\frac{5}{3}} & 0 \\
\frac{1}{\sqrt{3}} & -1 & -\frac{1}{\sqrt{15}} & - 2 \sqrt{\frac{2}{5}} \\
\sqrt{3} & 0 & 0 & 0
\end{array}
\right)
\eal
which acts on $|0, 1 \rangle$, $|1,1 \rangle$, $|2,1 \rangle$, and $|2, 3 \rangle$;  

\noindent
Even parity sector with $j=0$, 
%\chris{does not match (B22) of \cite{He:2023ewx} } 
%\sam{\cite{He:2023ewx} has been updated, now all the even sectors match, but the odd sectors have a minus sign (I got this result too)},
\bal
\frac{1}{\sqrt{3}} \left( 
\begin{array}{cc}
1 & - \sqrt{2} \\
-3 & 0 \\
\end{array}
\right)
\eal
which acts on $|0, 0 \rangle$ and $|2, 2 \rangle$; 

\noindent
Odd parity sector, $j \geq 2$,
\bal
\frac{(-1)^{j} 2^{\frac{j-3}{2}} (1+j)}{\sqrt{j (1+2j) {{2j \choose j}}}}
\left(
\begin{array}{cccc}
1 & \sqrt{\frac{j}{j+1}} & -\sqrt{\frac{j-1}{j+1}} & \sqrt{\frac{j+2}{j+1}} \\
1 & \sqrt{\frac{j}{j+1}} & \sqrt{\frac{j-1}{j+1}} & -\sqrt{\frac{j+2}{j+1}} \\
-\frac{2j +1}{j+1} & 0 & \frac{2j+1}{\sqrt{j^2-1}} & 0 \\
-\frac{1}{j+1} & -2 \sqrt{\frac{j}{j+1}} & \frac{3}{\sqrt{j^2-1}} & 2 \sqrt{\frac{j+2}{j+1}} 
\end{array}
\right)
\eal
which acts on $|1, j-1\rangle$, $|1, j+1 \rangle$, $|2, j-1 \rangle$, and $|2, j+1 \rangle$;  

\noindent
Odd parity sector, $j=1$,
\bal
\left(
\begin{array}{ccc}
-\frac{1}{\sqrt{6}} & -\frac{1}{2 \sqrt{3}} & -\frac{1}{2} \\
 -\frac{1}{\sqrt{6}} & - \frac{1}{2 \sqrt{3}} & \frac{1}{2} \\
- \frac{1}{\sqrt{6}} & \frac{1}{\sqrt{3}} & 0 
\end{array}
\right)
\eal
which acts on $|1,0 \rangle$, $|1, 2 \rangle$ and $|2, 2 \rangle$.  
For the odd parity sector $j=0$, we get $-\frac{1}{\sqrt{2}}$. This acts on $|1, 1 \rangle$.  

As a check, 
the parity of the state $|j_{12}, j_{123} \rangle$ 
is $(-1)^{j_1 - j_2 + j_3 - j_{123}} = (-1)^{j -j_{123}}$ (see (2.39) of \cite{Erramilli:2020rlr}), which 
agrees with the way the $| j_{12}, j_{123}  \rangle$ states have been
separated above. 

The conformal blocks output by Blocks\_3d are labeled by three SO(3) quantum numbers
 $(j_{12}, j_{120}, j_{34}, j_{340})$.   Our procedure for constructing the conserved
blocks is to contract a pair of the change of basis matrices above against a single conformal block and then contract again using a pair of OPE coefficient vectors.  After using the conservation conditions on the $\gamma^I$ and $\widetilde \gamma^I$,
the conserved conformal blocks have pairs of the remaining $\gamma^I$ and $\widetilde \gamma^I$ as coefficients.  
In the end, 
there are three types of conserved conformal blocks: even-even having two $\gamma^I$ coefficients multiplying them, even-odd having one $\gamma^I$ and one $\widetilde \gamma^I$, and odd-odd having two $\widetilde \gamma^I$ coefficients.
In the case where all the external operators are identical currents, one finds the following.
At $j=0$, there is one of each type of conserved block.  At $j=1$, Bose symmetry sets all the OPE coefficients to zero.
For $j\geq 3$ and odd, there is one odd-odd conformal block for each $j$.  For $j\geq 2$ and even, there is a single odd-odd conformal block, a 2$\times 1$ vector of even-odd blocks, and a $2 \times 2$ matrix of even-even blocks.
For a theory that preserves parity, all of the odd-even conformal blocks can be discarded.

\subsection{From the Conformal Blocks to the Crossing Equations}

In addition to the   $(j_{12}, j_{120}, j_{34}, j_{340})$ quantum numbers, the conformal blocks are labeled
by four azimuthal angular momenta $[q_1 q_2 q_3 q_4]$, 
with no constraint on the sum and $q_a \in \{ -j_a, -j_a+1, \ldots, j_a \}$.  
In what follows, we will focus primarily on these four numbers $[q_1 q_2 q_3 q_4]$.

The map from embedding and/or real 
space to these $q$-states is similar to what we saw above for the three-point function.
We will fix $x_1 = (0,0,0)$, $x_2 = (z, \bar z, 0)$, $x_3 = (0,1,0)$ and $x_4$ at infinity,
where we are using complex coordinates to parametrize the first two dimensions.  
More details can be found in Appendix F of \cite{Erramilli:2020rlr}.  
The basic idea is to decompose the four-point function into a sum over conformal blocks along with their 
tensor structures.
Our four-point function of four currents can be decomposed into conformal blocks
\begin{align}
	\expval{J J J J} \sim \sum_{\abs{q_i}\leq 1}[q_1 q_2 q_3 q_4]g_{[q_1 q_2 q_3 q_4]}(z,\overline{z}) \ .
\end{align}

Given that for the current four-point function $j_a = 1$, there are then 81 states. 
We take parity to be 
given by the action of the $\sigma_3$ Pauli matrix (instead of $\sigma_1$ as we did in the three-point function case)
which sends $\xi_a \to \xi_a$ and $\eta_a \to -\eta_a$.   
The parity is thus the parity of ${\sum_a (j_a - q_a)}$. There are 41 parity even and 40 parity odd states.  

For all identical currents, we can group the $q$-states together under orbits of the ${\mathbb Z}_2 \times {\mathbb Z}_2$ permutation group that leaves the cross ratios $z$ and $\bar z$ invariant.  There are 17 parity even orbits
\begin{align*}
& [0000] \\
& [1111] \\
& [{-}1{-}1{-}1{-}1] \\
& [0011], [1100] \\
& [0101], [1010] \\
& [0110], [1001] \\
& [00{-}1{-}1], [{-}1{-}100] \\
& [0{-}10{-}1], [{-}10{-}10] \\
& [0{-}1{-}10], [{-}100{-}1] \\
& [{-}1{-}111], [11{-}1{-}1] \\
& [{-}11{-}11], [1{-}11{-}1] \\
& [{-}111{-}1], [1{-}1{-}11] \\
& [111{-}1],[11{-}11],[1{-}111],[{-}1111] \\
& [{-}1{-}1{-}11],[{-}1{-}11{-}1],[{-}11{-}1{-}1],[1{-}1{-}1{-}1] \\
& [001{-}1],[00{-}11],[1{-}100],[{-}1100] \\
& [010{-}1],[10{-}10],[0{-}101],[{-}1010] \\
& [01{-}10],[100{-}1],[{-}1001],[0{-}110] 
\end{align*}
and 10 parity odd orbits
\begin{align*}
& [0001], [0010], [0100], [1000] \\
& [000{-}1],[00{-}10], [0{-}100], [{-}1000] \\
& [1110], [1101], [1011], [0111] \\
& [{-}1{-}1{-}10],[{-}1{-}10{-}1],[{-}10{-}1{-}1],[0{-}1{-}1{-}1] \\
& [110{-}1],[0{-}111],[11{-}10],[{-}1011] \\
& [011{-}1],[{-}1110],[10{-}11], [1{-}101] \\
& [01{-}11],[{-}1101],[101{-}1],[1{-}110] \\
& [{-}1{-}101],[01{-}1{-}1],[{-}1{-}110],[10{-}1{-}1] \\
& [0{-}1{-}11],[1{-}1{-}10],[{-}101{-}1], [{-}110{-}1] \\
& [0{-}11{-}1],[1{-}10{-}1],[{-}10{-}11],[{-}11{-}10]
\end{align*}
To be consistent with Bose symmetry, the conformal blocks assemble into permutation invariant combinations
\begin{eqnarray}
\langle q_1 q_2 q_3 q_4 \rangle &=& [ q_1 q_2 q_3 q_4] + n(1-z)^{-q_1+q_2+q_3-q_4} [q_2 q_1 q_4 q_3] 
+ n(z)^{q_1+q_2-q_3-q_4} [q_4 q_3 q_2 q_1] \nonumber \\
&&+ n(z)^{q_1+q_2-q_3-q_4} n(1-z)^{-q_1+q_2+q_3-q_4} [q_3 q_4 q_1 q_2] 
\end{eqnarray}
where $n(z) = \sqrt{z/\bar z}$. 
%\chris{double check signs}
The four-point function can be expressed then as 
\begin{align}
	\expval{J J J J} \sim \sum_{\abs{q_i}\leq 1} \langle q_1 q_2 q_3 q_4 \rangle g_{[q_1 q_2 q_3 q_4]}(z,\overline{z})
\end{align}
with the same conformal block coefficients as before.
This discussion can be repeated for $\langle JJ J'J' \rangle$ for which there is only a ${\mathbb Z}_2$ permutation symmetry,
25 even parity orbits and 20 odd parity orbits. 

 It remains to enforce current conservation.
Current conservation can be expressed in terms of the Todorov differential operator $D_w$ and the generator of rotations 
${\mathcal L}_{23}$ in the 23-plane:
\be
\left( \left(\frac{3}{2} - \omega \partial_\omega \right) \partial_{\bar \omega} \partial_z + 
\left(\frac{3}{2} - \bar \omega \partial_{\bar \omega} \right) \partial_\omega \partial_{\bar z}
+ \frac{i D_w^3 {\mathcal L}_{23}}{z- \bar z} \right) g(z, \bar z,  w_i) = 0 \ ,
\ee
\be
{\mathcal L}_{23} = i \sum_k \left( \omega_k^0 ( \partial_{\omega_k} - \partial_{\bar \omega_k}) + \frac{1}{2} (\omega_k - \bar \omega_k) \partial_{\omega^0_k} \right) \ . \nonumber
\ee
These expressions make use of the polarization notation
$\omega = w^z = w^1 + i w^2$, 
$\bar \omega = w^{\bar z} = w^1 - i w^2$,
$\omega^0 = w^3$.  The $q$-variables are the exponents of the complex variable $\omega$, where
$[1] = \omega$,  $[0] = \omega^0$, and $[-1] = \bar \omega$.

In the case of four identical currents and a parity preserving theory,  
the current conservation operator can be thought of as a map from a 17 dimensional subset of the full 81 dimensional 
$q$-state space to a smaller 
$q$-state space involving a scalar and three identical currents.  This 27 dimensional image space has
14 even parity states and 13 odd parity ones.
We followed a brute force approach and solved the conservation equation perturbatively in a series expansion near 
$z = \bar z = \frac{1}{2}$.  
The results depend sensitively on whether the currents are identical or not and whether or not the theory preserves parity.  

In the simplest case of four identical currents and where parity is preserved, we can find a solution if we provide initial data for five conformal blocks everywhere and two more conformal blocks ``along a line''.  More specifically, we must specify
the value of five conformal blocks and all their derivatives in $\partial_z$ and $\partial_{\bar z}$ at $z = \bar z = \frac{1}{2}$.  We must further specify for a sixth and seventh conformal block their values at $z = \bar z= \frac{1}{2}$ along with a subset of their derivatives, which we may 
take to be $\partial_z^n \partial_{\bar z}^{n-1}$ and  $\partial_z^{n} \partial_{\bar z}^{n}$ for all positive $n$.  

The point of this exercise is that it shows most of the 17 conformal blocks in this case are redundant and can be deduced from current conservation.  Thus we are well advised to construct crossing equations only from the remaining set.

This conservation constraint is sufficient when all the operators are identical.  However, 
to apply conservation to the $\langle J J J' J' \rangle$ four-point function, we can use the above expression on the 
second operator, but to ensure $J'$ is conserved as well, the strategy is first to perform the swap $1 \leftrightarrow 4$ and $2 \leftrightarrow 3$ and then act with the same conservation equation, which gives us a second set of constraints.

One last wrinkle is that the output of Blocks\_3d is actually linear combinations of the conformal blocks so far discussed:
\begin{equation}
g^\pm_{[q_1 q_2 q_3 q_4]} = \frac{1}{2} \left( g_{[q_1 q_2 q_3 q_4]} \pm (-1)^{\sum j_a}  g_{[-q_1 -q_2 -q_3 -q_4]} \right) \ .
\end{equation}

The basic crossing symmetry constraint is that 
\[
g_{[q_1 q_2 q_3 q_4]}(z, \bar z) = g_{[q_3 q_2 q_1 q_4]} (1-z, 1- \bar z) \ .
\]
Given this relation, we find the following cases when all the currents are identical.
We will leave an exposition of the more general case of two currents for part two of this project.

\subsubsection*{\texorpdfstring{$\langle JJJJ \rangle$}{<JJJJ>} Even Sector}

Conservation leaves 5 conformal blocks which are unconstrained, and two conformal blocks for which data needs to be provided just along a line.
One can consistently take $\langle 1111 \rangle$, $\langle 0110 \rangle$, $\langle 0101 \rangle$, 
$\langle 0011 \rangle$ and $\langle 0000 \rangle$ to be the five bulk blocks.  
$\langle{-}1{-}111 \rangle$ and $\langle{-}1 11 {-}1\rangle$ are a convenient choice for the line blocks, which
furthermore exchange under
$1 \leftrightarrow 3$.  
The crossing equations are
\begin{eqnarray}
\partial_z^n \partial_{\bar z}^{\bar n} g^+_{[1111]}(1/2, 1/2) &=& 0 \ , \; \; \; (n+\bar n \; \mbox{odd}) \ , \\
\partial_z^n \partial_{\bar z}^{\bar n} g^+_{[0101]}(1/2, 1/2) &=& 0 \ , \; \; \; (n+\bar n \; \mbox{odd}) \ , \\
\partial_z^n \partial_{\bar z}^{\bar n} g^+_{[0000]}(1/2, 1/2) &=& 0 \ , \; \; \; (n+\bar n \; \mbox{odd}) \ , \\
\partial_z^n \partial_{\bar z}^{\bar n} g^+_{[0011]}(1/2, 1/2) &=&  (-1)^{n+\bar n} 
\partial_z^n \partial_{\bar z}^{\bar n} g^+_{[1001]}(1/2, 1/2) \ .
\end{eqnarray}

\begin{eqnarray}
\partial_{\bar z}^n\partial_z^n g^+_{[-1-111]}(1/2, 1/2) &=&  \partial_{\bar z}^n \partial_z^n g^+_{[1-1-11]}(1/2, 1/2) \ , \\
\partial_{\bar z}^{n} \partial_z^{n+1} g^+_{[-1-111]}(1/2, 1/2) &=& 
 -\partial_{\bar z}^n \partial_z^{n+1} g^+_{[1-1-11]}(1/2, 1/2) \ .
\end{eqnarray}
Note that keeping purely holomorphic or antiholomorphic derivatives in the line constraint will fail to provide enough data to satisfy the conservation condition.

Curiously, while we find a line constraint, we were not able to implement successfully a line constraint in SDPB in all cases.  We found that for $\Lambda > 18$, we were unable to increase the maximal $n$ in the line constraint beyond 18.
There could be a reason we have not understood that makes the higher order line equations degenerate.  Indeed,
 our crossing equations are different from \cite{He:2023ewx}, where they find a point constraint instead of a line constraint.
In the numerics, we found we were able to get tighter bounds by working with a line constraint rather than a point constraint although the difference between the two decreases with increasing $\Lambda$, presumably because
the number of linear functionals from the five bulk constraints grows quadratically while the number of line constraint functionals grows only linearly.

\subsubsection*{\texorpdfstring{$\langle JJJJ \rangle$}{<JJJJ>} Odd Sector}

If we break parity, then we need to supply only two more conformal blocks.
$\langle 0111 \rangle$ and $\langle 0001 \rangle$ are a convenient choice. 
\begin{eqnarray}
\partial_z^n \partial_{\bar z}^{\bar n} g^+_{[1110]}(1/2, 1/2) &=& 0 \ , \; \; \; (n+\bar n \; \mbox{odd}) \ , \\
\partial_z^n \partial_{\bar z}^{\bar n} g^+_{[0001]}(1/2, 1/2) &=& 0 \ , \; \; \; (n+\bar n \; \mbox{odd}) \ .
\end{eqnarray}

\subsection{Parameters Used for SDPB and Blocks\_3d}

\vspace*{1em}
\begin{center}
\begin{tabular}{|l|c|}
\hline
%maxIterations & 500 \\
dualityGapThreshold & 1e-30 (1e-6 for OPE bounds) \\
primalErrorThreshold & 1e-30 \\
dualErrorThreshold & 1e-20 \\
initialMatrixScalePrimal & 1e20 \\
initialMatrixScaleDual & 1e20 \\
feasibleCenteringParameter & 0.1 \\
infeasibleCenteringParameter & 0.3 \\
stepLengthReduction & 0.7 \\
maxComplementarity & 1e100 \\
\hline
\end{tabular}

\vskip 0.1in

\begin{tabular}{|l|c|c|c|c|}
\hline
$\Lambda$ & 11 & 15 & 17 & 25 \\
precision & 650 & 650 & 650 & 1024 \\
$\ell_{\rm max}$ & 30 & 30 & 30 & 50 \\
{\rm kept\_pole\_order} & 20 & 20 & 20 & 20 \\
{\rm order} & 60 & 60 & 60 & 80 \\
\hline
\end{tabular}
\end{center}
\noindent
Despite some effort, we were never able to see dual jumps in the progress of SDPB, only primal jumps.  As a result, for the exclusion plots, an allowed point is one for which we observe a primal jump but a disallowed point corresponds to creeping behavior where both the dual and primal errors gradually creep down.  To distinguish these behaviors, we set the dualErrorThreshold higher than the primalErrorThreshold but still well below the values of primal error where the jumps were observed.

\section{Low-Lying Operators in 3d Free Theories}
\label{sec:free}

We describe the first few scalar primary operators and their correlation functions for the free three dimensional conformal field theories that make an appearance in this work: the generalized free vector field, the massless free scalar, and the free fermion.

\subsection{Generalized Free Vector Field}

The generalized free vector field $V^\mu(x)$ obeys the conservation condition $\partial_\mu V^\mu= 0$ and as a result 
has the two-point function $\langle V^\mu(x) V^\nu(0) \rangle = I^{\mu\nu}(x)/ |x|^4$.   Correlation functions all follow from Wick's Theorem.  
Writing a scalar operator as ${\mathcal O}_{\Delta, p}$ where $\Delta$ is its conformal dimension and $p$ the parity,
we are interested in the following three low-lying primary operators:
\begin{eqnarray}
\label{GFVFops}
{\mathcal O}_{4,+} &=& V^\mu V_\mu \ , \nonumber \\
{\mathcal O}_{5,-} &=& i \epsilon^{\mu\nu\lambda} V_\mu \partial_\nu V_\lambda \ , \\
{\mathcal O}_{7,-} &=& i \epsilon^{\mu\nu\lambda} \left[ 
(\partial_\rho V_\mu) (\partial^\rho \partial_\nu V_\lambda) - 4 (\partial^2 V_\mu) ( \partial_\nu V_\lambda) 
- V_\mu (\partial^2 \partial_\nu V_\lambda)
\right] \nonumber \ .
\end{eqnarray}
The operators ${\mathcal O}_{4,+}$ and ${\mathcal O}_{5,-}$ help us to identify the location of the GFVF near a cusp in the exclusion plot (fig.\ \ref{fig:exclusionplot}).
The dimension seven 
${\mathcal O}_{7,-}$
is the lowest odd parity scalar for the free scalar theory but is also present for the GFVF.

We find the following table of two and three-point function coefficients.  
The $\lambda_{VV\mathcal{O}}$ have been defined using the three-point
structures described in the text. 
\[
\begin{array}{c|ccc}
& c_{\mathcal{OO}} & \gamma_{VV \mathcal{O}} &\lambda^2_{VV \mathcal{O}} \\
\hline
{\mathcal O}_{4,+} & 6 & 1 & \frac{1}{6} \\
{\mathcal O}_{5,-} & 12 & 2 & \frac{2}{3} \\
{\mathcal O}_{7,-} & 12{,}960 & 36 & \frac{1}{5} 
\end{array}
\]
The normalized OPE coefficient is defined such that
\[
\lambda^2_{VV\mathcal{O}} \equiv \frac{\gamma^2_{VV\mathcal{O}}}{\tau^2_{VV} c_{\mathcal{OO}}} \ .
\]
The results for ${\mathcal O}_{4,+}$ are consistent with the results discussed for the free Maxwell field, where 
${\mathcal O}_{4,+}$ was identified with the displacement operator $D$, and $V^\mu$ with the boundary current $B^\mu$.  With the normalizations used there, 
we found that $C_D = \frac{6}{\pi^4}$, the current two-point function was normalized as
$\tau_{BB} = \frac{2 g^2}{\pi^2}$, and the three-point function is $\gamma_{BBD}= \frac{2 g^2}{\pi^4}$.

The result $\lambda_{VVO}^2 = \frac{1}{6}$ for ${\mathcal O}_{4,+}$ 
was crucial for
ensuring the proper normalization of the OPE coefficient bounds in the text.  The values 
for the odd parity operators 
we intend to make use of future work.

\subsection{Free Scalar}

Using the two-point function
\[
\langle \phi(x) \phi(0) \rangle = \frac{1}{|x|} \ ,
\]
we can compute the OPE coefficients of some low lying primary operators in the spectrum.  
In particular, we are interested in the current
\[
J_\mu = i (\phi \partial_\mu \bar \phi - \bar \phi \partial_\mu \phi)  \ .
\]
With this choice of normalization,  $\langle J_\mu(x) J_\nu(0) \rangle = 2 I_{\mu\nu}(x) / |x|^4$.  
The lightest even parity  scalar is ${\mathcal O}_{1,+} = |\phi|^2$.
The next even parity scalar has dimension four.  Note that $|\partial_\mu \phi|^2$ is a descendant operator, 
while neither $|\phi|^4$ nor $|\phi|^6$ are present in the OPE of two currents in the free theory.

The lightest odd parity scalar can be deduced from 
the result (\ref{GFVFops}) for ${\mathcal O}_{7,-}$ above along with the equation of motion $\Box \phi = 0$:
\[
{\mathcal O}_{7,-} = i \epsilon^{\mu\nu\lambda} \left[2 J_\mu (\partial^\rho \partial_\nu \phi )(\partial_\rho \partial_\lambda \bar \phi) 
+3(\partial^2 J_\mu)  ( \partial_\nu \phi )(\partial_\lambda \bar \phi) \right] \ .
\]
The dimension five operator $ \epsilon J \partial J$ that we found for a GFVF vanishes for a single complex scalar field.  Writing such an operator for free scalar fields 
requires at least two commuting currents, $J = \frac{1}{\sqrt{2}} (J_1 + J_2)$.  
We see that the exclusion plot (fig.\ \ref{fig:exclusionplot}) bends upward at small $\Delta_+$ to allow for an 
${\mathcal O}_{7,-}$ operator.

We find the following table of two and three-point function coefficients:
\[
\begin{array}{c|ccc}
& c_{\mathcal{OO}} & \gamma_{JJ \mathcal{O}} & \lambda^2_{JJ \mathcal{O}} \\
\hline
{\mathcal O}_{1,+} & 1 & 2 &  1 \\
{\mathcal O}_{7,-} & 8640 &  72 & \frac{3}{10} 
\end{array}
\]

\subsection{Free Fermion}

Our building block is the two-point function of two fermionic operators
\[
\langle \psi(x) \bar \psi(0) \rangle = \frac{\gamma_\mu x^\mu}{|x|^3} \ .
\]
For the current operator $J^\mu = \bar \psi \gamma^\mu \psi$, the two-point function is
the same as for the free scalar, 
$\langle J^\mu(x) J^\nu(0) \rangle = 2 I^{\mu\nu}(x) / |x|^4$.  
The mass breaks parity in three dimensions, and correspondingly the operator ${\mathcal O}_{2,-} = \bar \psi \psi$
is parity odd.  
There is an even parity dimension four operator ${\mathcal O}_{4,+} = (\bar \psi \psi)^2$.  By a Fierz identity, 
$(\bar \psi \psi)^2 \sim  J^\mu J_\mu$.
The two operators ${\mathcal O}_{4,+}$ and ${\mathcal O}_{2,-}$ lie safely inside the exclusion plot (fig.\ \ref{fig:exclusionplot}).
The value of $\lambda^2_{JJO}$ for the ${\mathcal O}_{4,+}$
operator was recovered as an upper bound on an OPE coefficient in fig.\ \ref{fig:ratioOPEplot}.
\[
\begin{array}{c|ccc}
& c_{\mathcal{OO}} & \gamma_{JJ \mathcal{O}} & \lambda^2_{JJ \mathcal{O}} \\
\hline
{\mathcal O}_{4,+} & 4  & 2  & \frac{1}{4} \\
{\mathcal O}_{2,-} & 2 & 4 &  2  % \\
%{\mathcal O}_{7,-} &  &   & 
\end{array}
\]

\section{Current-Current-Displacement Three-Point Function}
\label{sec:Ward}

While three-point functions in conformal field theory are normally fixed up to constants, 
the current-current-displacement three-point function in our set-up is special.  It can be expressed in terms
of the two-point functions of the currents and the displacement operator.  
As the relation is important for us, we reproduce here the derivation of ref.\  \cite{DiPietro:2019hqe}.

Expanding $\langle F_{\mu\nu} F_{\lambda\rho} D \rangle$ using the BOE (\ref{eq:Fope}), we can express this bulk-bulk-boundary three-point function in terms of the following purely boundary correlation functions:
\begin{align}
\langle F_{in}(x) F_{jn}(y) D(z) \rangle &= \langle E_i(x) E_j(y) D(z) \rangle + \ldots 
\ , \\
\langle F_{in}(x) F_{jk}(y) D(z) \rangle &= i \langle E_i(x) B^l(y) D(z) \rangle \epsilon_{jkl} + \ldots 
\ , \\
\langle F_{ij}(x) F_{kl}(y) D(z) \rangle &= - \langle B^m(x) B^n(y) D(z) \rangle \epsilon_{mij} \epsilon_{nkl} + \ldots
\ ,
\end{align}
where the ellipses denote higher order terms in the distance from the boundary.

Conformal invariance constrains current-current-scalar three-point functions to take the form
\begin{align}
	\expval{J_{(a)}(x,z_1)J_{(b)}(y,z_2)D(z)}&=\frac{\gamma_{ab}^{1}H_{12}+\gamma_{ab}^{2}V_1 V_2 + \widetilde{\gamma}_{ab}^{1}\epsilon_{12}}{(x-z)^{4}(y-z)^4} \ .
\end{align}
The structures are defined in section \ref{sec:threepoint}. Conservation of $J_{(a)}$ imposes $\widetilde{\gamma}_{ab}^{1}=0$, and $\gamma_{ab}^{2}=2\gamma_{ab}^{1}$. 
Therefore we find that
\begin{align}
\lim_{|z|\to \infty} |z|^8 \langle F_{in}(x) F_{jn}(y) D(z) \rangle &=\gamma_{EE}^1 \delta_{ij} + \ldots 
\ , \\
\lim_{|z|\to \infty} |z|^8 \langle F_{in}(x) F_{jk}(y) D(z) \rangle &= i  \gamma_{EB}^1\epsilon_{ijk} + \ldots 
\ , \\
\lim_{|z|\to \infty} |z|^8 \langle F_{ij}(x) F_{kl}(y) D(z) \rangle &= - \gamma_{BB}^1 (\delta_{ik} \delta_{jl} - \delta_{il} \delta_{kj}) + \ldots
\ .
\end{align}

We compare this boundary limit with the coincident limit $x \to y$.  The bulk OPE of $F_{\mu\nu}$ with itself is fixed by the free theory, and there are only three operators in this OPE that will 
contribute to the $\langle F F D \rangle$ three-point function.  They are $F^2$, $F \widetilde F$, and the stress tensor 
$T_{\mu\nu} = \frac{1}{g^2} (F_{\mu\rho} {F_{\nu}}^{\rho} - \frac{1}{4} \delta_{ab} F^2)$ which come with coefficients
\begin{align}
F_{\mu\nu}(x) F^{\lambda \rho} (y) \sim \frac{1}{12} (\delta_\mu^\lambda \delta_\nu^\rho - \delta_\mu^\rho \delta_\nu^ \lambda) {:}F^2(y){:} + \frac{1}{12} {\epsilon_{\mu\nu}}^{\lambda \rho} {:}F \widetilde F(y){:} + 2 g^2 \delta^{[\lambda}_{[\mu} T^{\rho]}_{\nu]}
\ . 
\end{align}
Note other higher spin operators are present in the OPE of $F_{\mu\nu}$ with itself but they have zero overlap with
the displacement operator.  It is precisely this point which makes the calculation useful; in the more general case, there would be an infinite sum over bulk operators over which we would have less fine control.

By the constraints of conformal invariance, the following bulk-boundary two-point functions are all fixed up to constants:
\begin{align}
\langle {:} F^2(x) {:} D(z) \rangle &= \frac{b_{F^2 ,D}}{|x-z|^8} \ , \\
 \langle {:} F\widetilde F(x) {:} D(z) \rangle &= \frac{b_{F\widetilde F ,D}}{|x-z|^8} \ , \\
\langle T_{\mu\nu}(x) D(z) \rangle &= \frac{b_{T,D} \left( X_\mu X_\nu - \frac{1}{4} \delta_{\mu\nu} \right)}{|x-z|^8} \ .
\end{align}
That the boundary limit of $T_{nn} = D$ tells us that 
\begin{align}
b_{T,D } = \frac{4}{3} C_D
\end{align}
where $C_D$ is the coefficient of the displacement two-point function.
We can then use the Ward identity that for a general bulk operator ${\mathcal O}$
\be
\int d^{d-1} y \langle {\mathcal O}(x) D(y) \rangle = \partial_n \langle {\mathcal O}(x) \rangle \ , 
\ee
which yields from (\ref{eq:FFonepoint}) the following relations
\begin{align}
b_{F^2 ,D} = - \frac{12 \kappa \chi_e a_{F^2}}{\pi^2} \ , \; \; \; b_{F \widetilde F ,D} = - \frac{12 \kappa \chi_o a_{F \widetilde F}}{\pi^2} \ .
\end{align}

Assembling the chain of relations, we find that
\bal
	\gamma_{EE}^{1}&= - \frac{\kappa \chi_e}{\pi^2}  + \frac{g^2 C_D}{3}\ , \\
	\gamma_{EB}^{1}&= i \frac{\kappa \chi_o}{\pi^2} \ , \\
	\gamma_{BB}^{1}&=   \frac{\kappa \chi_e}{\pi^2}+ \frac{g^2 C_D}{3} \ ,
\eal
which can be recast into the form
\bal
\label{gammaJJD}
	\gamma_{EE}^{1}&= \frac{2}{\pi^2}\tau_{EE} +\frac{\kappa}{\pi^2}\left(\frac{C_{D}}{C_{D}^{free}}-1\right) \ , \\
	\gamma_{EB}^{1}&= \frac{2}{\pi^2} \tau_{EB} \ , \\
	\gamma_{BB}^{1}&= \frac{2}{\pi^2} \tau_{BB} +\frac{\kappa}{\pi^2}\left(\frac{C_{D}}{C_{D}^{free}}-1\right) \ ,
\eal
using that $\tau_{EE} + \tau_{BB} = \kappa$, the definitions (\ref{eq:crels}) and (\ref{eq:kappachidefs}), and the result (\ref{eq:DDrel}) for $C_D^{free}$.  
It would be interesting to look at a more general version of this calculation where $D$ is replaced in the three-point function with the boundary value of a higher spin primary in the bulk OPE of two Maxwell fields.

\bibliographystyle{JHEP}
\bibliography{bib}

\end{document}